\documentclass[twoside]{article}  
\usepackage{graphicx}
\usepackage{txfonts}
\usepackage{natbib}
\usepackage{longtable}
\usepackage{xcolor}
\usepackage{psfig}

\begin{document}
\noindent
{\hfill{9 February 2023}}
\begin{center}
\vskip 0.5 cm
{\Large \bf Atlas of photometric and spectroscopic daily monitoring
of the very slow}\\ 
\smallskip
{\Large \bf 
Nova Cas 2021 (=V1405 Cas) during the first 660 days of its outburst} \\
\vskip 0.7cm
{\large P.~Valisa$^1$, U.~Munari$^2$, S.~Dallaporta$^1$, A.~Maitan$^1$, and A.~Vagnozzi$^1$ }
\end{center}
\vskip 0.3 cm
\noindent
\phantom{~~~~~~~~~~~}{\small 1: ANS Collaboration, c/o Astronomical Observatory, 36012 Asiago, Italy}\\
\phantom{~~~~~~~~~~~}{\small 2: INAF Astronomical Observatory of Padova, 36012 Asiago, Italy}\\
\vskip 0.7cm

\baselineskip 7pt
\begin{center}
\parbox{14.5cm}{
{\bf Abstract}. {\footnotesize
Nova Cas 2021 erupted on 18 March 2021, reaching naked-eye brightness when
passing through photometric maximum 53 days later.  The fact that the nova
has been declining very slowly and it is circumpolar from our observing
locations, allowed us to monitor its evolution at $\sim$daily temporal
cadence and without seasonal gaps, covering the first 660 days since
discovery.  In all we obtained 574 highly accurate photometric runs
simultaneously in the $B$$V$$R$$I$ bands, which were distributed over
313 different nights, and 110 Echelle high-resolution, high S/N spectra were
recorded over 107 individual nights.

The multi-band photometric evolution is mapped in detail, and the strict
similarities to the proto-type very-slow novae HR~Del and V723~Cas are
discussed.  In addition to an initial extended plateau, lasting for several
months, all three novae displayed multiple and short-lasting superimposed
maxima, the first being invariably the brightest, and the last the widest
and leading directly into the decline and the nebular phase, which progressed 
along an increasing ionization degree up to the appearance of coronal lines years
past maximum.  The decline proceeded quite smoothly for all three novae,
with an identical exponential decline of the $V$-band flux: $F^{V}_{decline} \propto
(t-t_0)^{\alpha}$, with $\alpha$=$-$2.33, $-$2.34, and $-$2.29 for Nova Cas
2021, V723 Cas, and HR Del, respectively.  At the current decline rate, Nova
Cas 2021 will return to within 10\% of the pre-outburst brightness level
only by the spring of 2029.

Nova Cas 2021 belongs to the spectroscopic FeII-class, with continuous changes
of line intensity and profiles in response to the reckless photometric
activity during the plateau, which lasted for the initial seven months
following the nova discovery.  The immense amount of information stored in
our high-resolution Echelle spectra covering the whole optical range up to
OI 8446 and the CaI triplet, suggested us to develop 2D-dynamic
presentations of the most sensitive spectral ranges, to highlight the
temporal evolution of the multi-component P-Cyg profiles, the intensity and
profile of the many emission lines, and the ionization/excitation response
to the photometric changes.

P-Cyg absorptions decelerated (from $-$1500 to $-$800 km/s) during the
initial 53-days leading up to the primary maximum, remained stable although
multi-component during the main part of the plateau phase (days +65 to +140),
and then progressively accelerated up to $-$2200 km/s during early decline from
plateau, finally disappearing completely by day +330.  Upon leaving the
plateau, nebular lines appeared and the ionization degree quickly
increased, passing from FeII/Balmer/HeI through HeII/Bowen and then to
[CaV]/[NeV]/[FeVII].  At day +662 [FeX] has not yet appeared, but this could
happen soon considering the continuing growth in intensity of [FeVII],
which on day +662 is second in intensity only to [NeV] and H$\alpha$.  

Prior to primary maximum, the emission lines were characterized by a slim Voigt
profile, afterward their width flared-up with a reckless variability at all
velocity scales in response to the ever changing brightness of the nova
during the plateau and the secondary maxima.  After about day +550, the profiles
stopped to evolve, freezing their configuration best described as a
trapezoidal pedestal, 2620~km/s wide at the base and 1930~km/s at the top,
with superimposed a central core 930~km/s wide at the base and 800~km/s at the
top (values for [FeVII] on day +662).  At later epochs, the profiles of all
emission lines turned densely castellated: no change larger than 1 km/s has
been measured for any dent presented by [FeVII] during the last recorded 110
days.

The passage at photometric maximum on day +53 changed irreversibly the
structure of emission lines: with the return to plateau brightness of the
nova on day +62, all lines became much wider, developing a broad pedestal
and a superimposed wide central core, indicative of separated and fast
velocity outflows.  It is catching that this happened simultaneous with the
detection of shock-induced $\gamma$-rays from Nova Cas 2021, which lasted
for five days until day +66.

}}
\end{center} 
\baselineskip 13pt

\noindent
\section{Introduction}
\bigskip

Nova Cas 2021 (= V1405 Cas; {\bf NCas21} for short) was discovered at
unfiltered 9.6mag on 2021 March 18.424 UT by Yuji Nakamura (Japan) on CCD
frames taken through a 135mm/F4.0 lens, and was logged by CBAT as transient
PNV J23244760+6111140.  The object was soon classified as a classical nova
by Maehara et al.  (2021) based on an optical spectrum obtained within a few
hours of discovery on March 18.820 UT.

Preliminary reports on the photometric and spectroscopic appearance of NCas21 at
optical and IR wavelengths during the 7-months long plateau around maximum
brightness were issued by Taguchi et al.  (2021a,b), Munari et al. 
(2021a,b), Gehrz et al.  (2021), Rudy et al.  (2021), Shore et al. 
(2021a,b), and Woodward et al.  (2021), with Munari et al.  (2021c) finally
reporting in early December 2021 the start of the photometric decline and the
simultaneous turn toward high ionization conditions of the optical spectra. 
The latter prompted new {\it Swift} observations by Page et al.  (2021) that
revealed the emergence of a faint super-soft emission component at X-rays,
which was absent in earlier {\it NuSTAR} and {\it Swift} observations that
detected only weak and much harder X-rays from NCas21 (Sokolovsky et al. 
2021a).

Radio observations of NCas21 with VLA by Sokolovsky et al.  (2021b, 2022a) revealed
strong emission at the time the nova begun the photometric decline in
December 2021, with a spectral index $\alpha$=+1.86$\pm$0.02 indicative of 
thermal and optically thick emission, as later confirmed by
the uGMRT observations of Nayana et al.  (2022).  The radio emission of the
nova was then spatially resolved on VLA observations obtained in July 2022
by Sokolovsky et al.  (2022b), in the form of an elliptical edge-brightened
shell extending about 220 mas in the north-east to south-west direction
crossed by a bright band of emission extending about 120 mas along the minor
axis of the shell.

NCas21 has been recently classified among the exceedingly rare class of {\it
neon} novae by Munari \& Valisa (2022), based on the exceptional intensity
reached by [NeV] 3346, 3427 \AA\ on their spectra for day +618, by far at that time 
the strongest emission lines over the whole near-UV/optical range.

In this paper we present a quick overview of the BVRI photometric and Echelle
spectroscopic results we gathered during our $\sim$daily monitoring of NCas21
that covers - with no seasonal gap - the first 660 days of the outburst.
  
\section{BVRI Photometry}

\subsection{Observations}

$B$$V$$R$$I$ optical photometry of NCas21 on the Landolt (2009) photometric
system has been collected daily by ANS Collaboration, primarily with
telescopes ID 0310, 1301, and 2203, all adopting Astrodon photometric
filters of a recent generation correcting for the red-leak affecting older
$B$-filter versions.  In all 574 $B$$V$$R$$I$ runs spread over 313 different nights
were collected.  The fact that all telescopes adopt the same set of filters
helps in suppressing the unavoidable scatter of the measurements induced by
differences in the actual pass-bands, when dealing with objects
characterized by the presence of very strong emission lines in their
spectra.

Technical details and operational procedures of the ANS Collaboration
network of telescopes running since 2005, are presented by Munari et al. 
(2012).  Detailed analysis of the photometric performances and multi-epoch
measurements of the actual transmission profiles for all the photometric
filter sets in use at all telescopes is presented by Munari \& Moretti
(2012), and updated in Munari et al.  (2015).  The same local photometric
sequence, extracted from APASS DR8 (cf.  Henden \& Munari 2014, Munari et
al.  2014) and spanning a wide color range, has been used at all telescopes
on all observing epochs, ensuing a high consistency of the data which have
been transformed from the local instantaneous to the standard Landolt system
via color-equations.  All measurements were carried out with aperture
photometry, with the radius of the aperture generally about 1.0 FWHM and
optimized on each observations so to minimize the dispersion of the
comparison sequence around the transformation color relations.  Colors and
magnitudes are obtained separately during the reduction process, and are not
derived one from the other.  The total error budget, which quadratically
combines the measurement error on the variable with the error associated to
the transformation from the local instantaneous to the standard photometric
system (as defined by the photometric comparison sequence), are $<$0.01 mag
for all NCas21 observations in all bands.

	\begin{figure}[!hb]
	\centering
	\includegraphics[width=15.3cm]{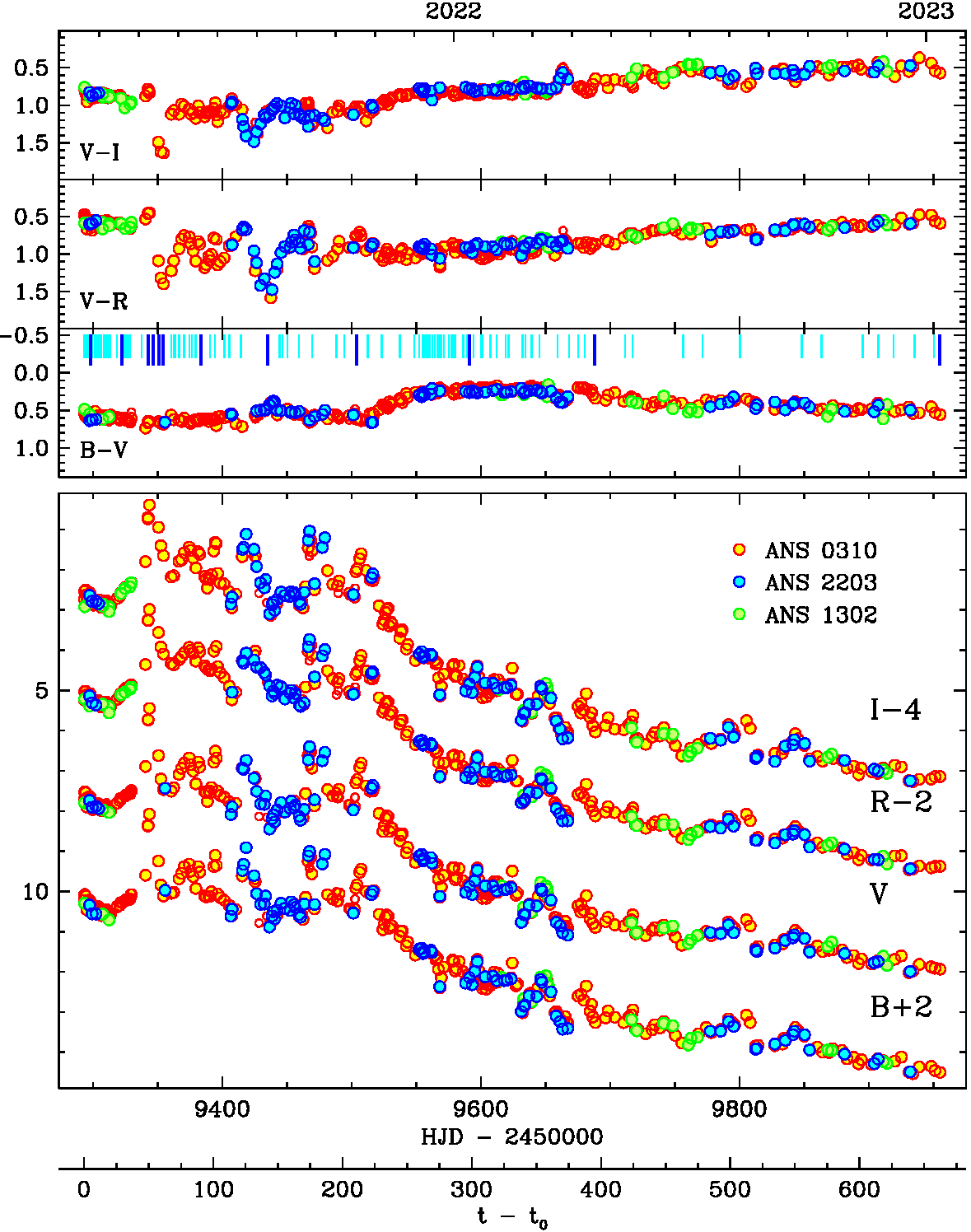}
	\caption{ANS Collaboration $B$$V$$R$$I$ color- and light-curves of Nova Cas
	2021.  The observing epochs are counted as days passed since
	discovery (cf.  Eq.(1)).  The thin cyan ticks in the
	$(B-V)$ panel mark the epoch of Echelle observations, the thicker
	and blue ticks points to the epoch of Echelle spectra plotted in
	Figure~\ref{fig_12_date}.}
	\label{fig_BVRI}
	\end{figure}

\bigskip
\noindent
\subsection{Photometric evolution}
\bigskip

The ANS Collaboration light- and color-curves covering the first 660 days of
NCas21 are presented in Figure~\ref{fig_BVRI}.  In this figure and in the rest of
this paper, the $t_0$ reference epoch is taken to coincide with the
discovery of the nova on:
\begin{equation}
t_0 = 2459292.924~{\rm HJD}  ~~~~~~ (= 2021 ~{\rm March}~ 18.424~{\rm UT})   
\end{equation}

The nova spent the first $\sim$7 months lingering around a bright plateau, characterized by a series of 
isolated maxima of varied height and width.  The epoch, width, and 
peak brightness of the five main maxima are listed in Table~\ref{tab_maxima}. 
The \underline{median} values during the 7-months plateau have been:
\begin{equation}
B_{\rm plateau} = 8.848 ~~~~~ V_{\rm plateau} = 8.407 ~~~~~ R_{\rm plateau} 
= 7.732 ~~~~~ I_{\rm plateau} = 7.401 ~~ {\rm mag}
\end{equation}
The end of the plateau is taken to be 5 Nov 2021, day +230, when the nova
crossed these median values while declining from the last re-brightening (N.5
in Table~\ref{tab_maxima}) and embracing the final, steady decline.  The maximum brightness
(N.1 in Table~\ref{tab_maxima}) occurred on:
\begin{equation}
{\rm max:}~~~~~~~~~~~~{\rm HJD} 2459345.9\pm0.3~~=~~2021~{\rm May}~11.4~{\rm UT}~~~~~~~~ ({\rm day}~+53.0) 
\end{equation}
and has been characterized by 
\begin{equation}
V_{\rm max} = 5.35 ~~~~~ (B-V)_{\rm max} = +0.662 ~~~~~ (V-R)_{\rm max} 
= +0.452 ~~~~~ (V-I)_{\rm max} = +0.833
\end{equation}

The MMRD relation (magnitude at maximum vs.  rated of decline) is frequently
used to estimated the distance to a nova.  Its most recent formulation by
Selvelli \& Gilmozzi (2019), based on Gaia DR2 parallaxes, makes use of
$t_3$ as the time the nova takes to decline by 3.0 mag below the peak
$V$-band brightness.  Unfortunately, a light-curve as complex as that of
NCas21, exhibits multiple crossing of the $\Delta V$=3.0mag threshold, with
different resulting absolute magnitudes and derived distances.  These
different possible value of $t_3$ for NCas21 are listed in
Table~\ref{tab_t3}, which also reports the corresponding absolute magnitude
following Selvelli \& Gilmozzi (2019), and the distances resulting from
adopting the reddening $E_{B-V}$=0.50 derived below in Sect.~3.3.

As noted by Taguchi et al.  (2021a), NCas21 is coincident with the
astrometric position of the faint and blue variable CzeV3217, characterized
by pre-outburst PanSTARRS PS1 mean magnitudes of $g'$=15.72, $r'$=15.54,
$i'$=15.50, $z'$=15.36, $Y$=15.43 and having a photometric period of
0.1883907 days = 4h 31min.  The outburst amplitude of $\Delta$mag$\sim$10.5
is in excellent agreement with the Warner (1995, his Figure 5.4) relation
between amplitude and time to decline from maximum, supporting CzeV3217 as
the true progenitor of NCas21.  Gaia DR3 provides a robust parallax of
0.5776$\pm$0.0254 mas for CzeV3217, corresponding to a distance of 1.73
(range 1.66$\leftrightarrow$1.81) kpc, which is in reasonable agreement with the
distance to NCas21 derived for $t_3$=90 days in Table~\ref{tab_t3}. At this distance,
the ellipsoidal radio structure resolved by Sokolovsky et
al. (2022b) with VLA is expanding on the plane of the sky along its major axis 
at an averaged velocity of:
\begin{equation}
v_\perp ~=~ 700 ~~~{\rm km/s} 
\end{equation}

	\begin{table}
	\centering
	\makebox[0pt][c]{\parbox{1.0\textwidth}{%
	    \begin{minipage}[t]{0.42\hsize}\centering
		\caption{Epoch, width, and peak $V$-band brightness of the five
	        highest maxima displayed by Nova Cas 2021 during the plateau.} 
		\begin{tabular}{ccrrc}
		&&\\
		\hline
	        &&\\
		N.  & HJD & $t-t_0$ & FWHM & $V_{peak}$ \\
	            &(-2450000)&(days)&(days)&(mag)\\
		&&\\
		1   & 9345.9 & 53.0  &  6.8 & 5.35      \\
	        2   & 9420.2 & 127.3 & 10.5 & 6.09      \\
		3   & 9467.8 & 174.8 &  3.1 & 6.40      \\
	        4   & 9478.8 & 185.9 &  5.9 & 6.55      \\
	        5   & 9508.0 & 215.0 & 12.3 & 6.80      \\
		&&\\
		\hline
		\end{tabular}
	        \label{tab_maxima}
	    \end{minipage}
	    \hfill
	    \begin{minipage}[t]{0.50\hsize}\centering
	            \caption{Epochs past primary maximum (the N.1 in Table~\ref{tab_maxima})
                     when Nova Cas 2021 crossed the $\Delta V$=3.0mag
	             threshold below the peak brightness ($t_3$), and the corresponding $M_V$
	             and distances.}
	            \begin{tabular}{rcl}
	            &&\\
	            \hline
	             &&\\
	             $t_3$  & $M_V$ & dist \\
	             (days) &(mag)  & (kpc)\\
	            &&\\
	            90  & $-$6.9 & 1.4 \\
	             154 & $-$6.4 & 1.1 \\
	             180 & $-$6.3 & 1.0 \\
	             266 & $-$5.9 & 0.85 \\
	            &&\\
	            \hline
	            \end{tabular}
	        \label{tab_t3}
	    \end{minipage}%
	}}
	\end{table}

The evolution of NCas21 during the plateau is markedly different among
different optical colors.  It has been characterized by a flat behavior in
$(B-V)$, passing basically unchanged through all the various re-brightening,
while $(V-R)$ and $(R-I)$ varied in anti-phase by more than one magnitude,
and much less for $(V-I)$, which points to the $R$ band as main culprit for
the observed changes in color.  When NCas21 left the plateau and begun the
steady decline, all colors reacted by changing their trend and smoothing out
any short-term variability, a fact that is most evident for $(B-V)$ in
Figure~\ref{fig_BVRI}.

The post-plateau flux radiated by NCas21 through the $V$-band is reducing
exponentially as
\begin{equation}
F^{V}_{decline} \propto (t-t_0)^{-2.33}
\end{equation}

At this rate, NCas21 will have declined within 10\% of the pre-outburst
level by the spring of 2029.

\subsection{Periodic and rapid variability during the decline}

\begin{figure}
	 \begin{minipage}[b]{8.1cm}
	   \centering
	   \includegraphics[width=5.9cm]{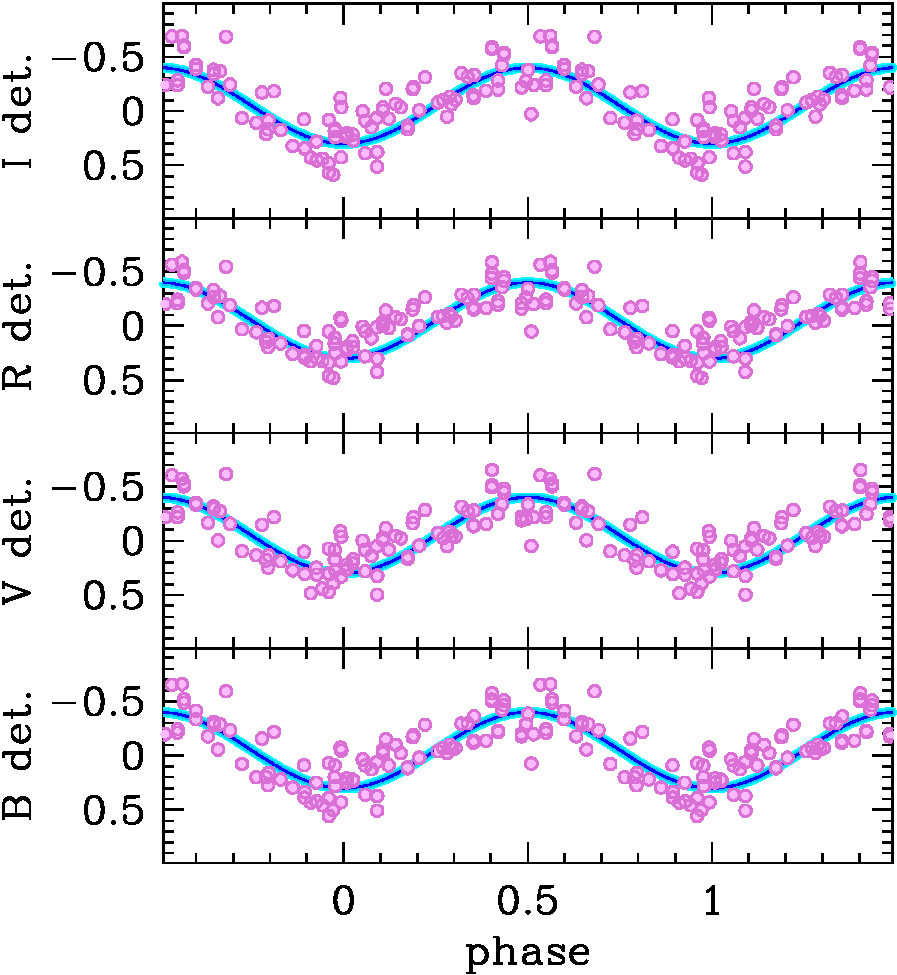}
	   \caption{De-trended Nova Cas 2021 data from Figure~\ref{fig_BVRI}
	   for +300$\leq (t-t_0)\leq$+400, are phase-plotted against the
	   P=30.5 days ephemeris of Eq.(7).}
	\label{fig_period}
	 \end{minipage}
	 \ \hspace{2mm} \hspace{3mm} \
	 \begin{minipage}[b]{8.1cm}
	  \centering
	   \includegraphics[width=5.9cm]{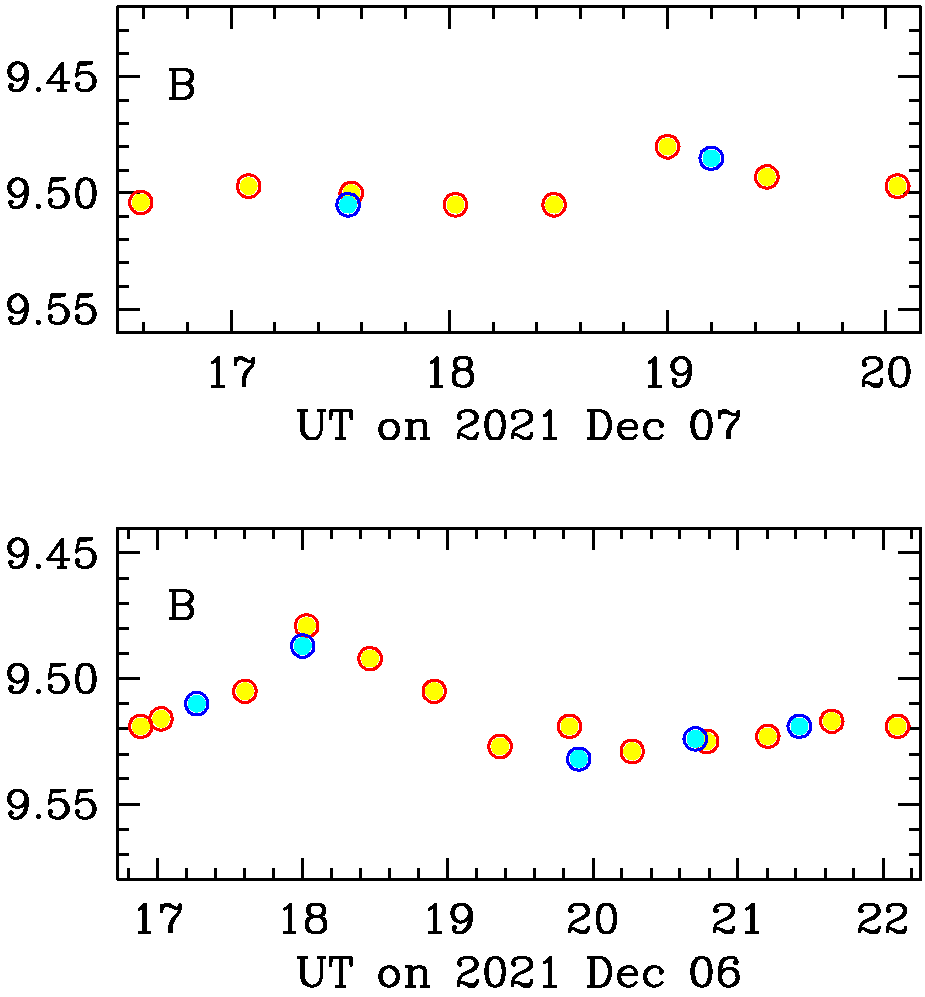}
	   \caption{Data from Figure~\ref{fig_BVRI} for 2021 Dec 6 and 7
	   (day +262 and +263) are plotted on an expanded scale to highlight
	   the level of intra-night variability affecting the early decline 
           of Nova Cas 2021.}
	\label{fig_intranight}
	 \end{minipage}
	\end{figure}

During the otherwise rather smooth exponential decline, NCas21 exhibited a
periodic variability lasting a few consecutive cycles between about days
+300 and +400, in the form of a sinusoidal wave of 0.7mag total amplitude
and the ephemeris for minima given by:
\begin{equation}
{\rm min} = 2459694 + 30.5\times E
\end{equation}
The corresponding phased light-curve of NCas21 is presented in
Figure~\ref{fig_period}, after removing the underlying exponential trend of
the decline.  All bands behaved rather similarly, and an equivalent plot for
the photometric colors confirms that the large amplitude sinusoidal
variability was not accompanied by significant changes in the spectral
energy distribution of NCas21.

A Fourier analysis of the de-trended photometry of NCas21 during the decline
phase does not reveal any significant periodicity in addition to P=30.5
days, the latter being confined to the day +300 to +400 time interval and
not present before or after.

In particular no significant Fourier peak appears associated with the
P=0.1883907 days variability exhibited by the progenitor of the nova.  There
is a lot of intra-night variability, as for ex.  plotted in
Fig~\ref{fig_intranight} for 2021 Dec 6, but such variability looks rather
chaotic.  In fact, similar data for the next night looked rather flat.

\subsection{Comparison with the prototype very-slow novae HR Del and V723 Cas}

	\begin{figure}[!ht]
	\centering
	\includegraphics[width=16cm]{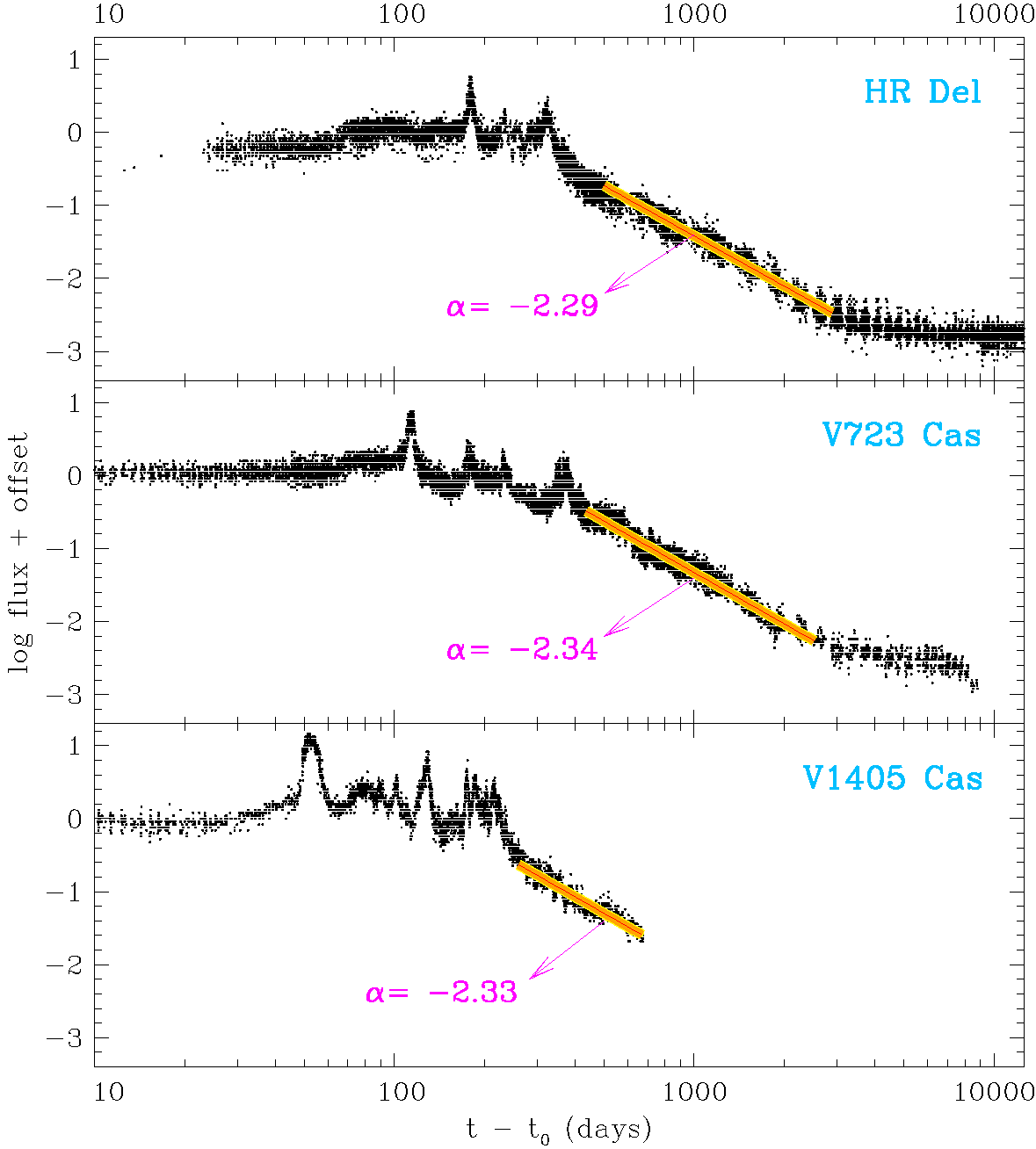}
	\caption{The evolution of Nova Cas 2021 and of prototype slow-novae
	HR Del (1967) and V723 Cas (1995) are compared on a {\it log-log}
	plot between the time elapsed since discovery and the flux radiated
	through the $V$ passband, normalized to median value during the
	plateau.  The decline from plateau is fitted with an exponential
	relation of the type $F^{V}_{decline}$$\propto$(t-t$_0)^\alpha$ and
	the resulting value of the $\alpha$ index is given.  Input data: visual
	estimates collected by AAVSO.}
	\label{fig_3novae}
	\end{figure}

The extended plateau at maximum, the several superimposed re-brightening,
and the very slow decline exhibited so far by NCas21 are the hallmarks of
the prototype slow-novae HR Del (1967) and V723 Cas (1995), which main
properties have been discussed - among others - by Rafanelli \& Rosino
(1978), Bruch (1982), Munari et al.  (1996), Harman \& O'Brien (2003),
Heywood et al.  (2005), Iijima (2006), Goranskij et al.  (2007), Lyke \&
Campbell (2009), Schwarz et al.  (2011), Ochner et al.  (2015), and Chomiuk
et al.  (2021).

The light-curves of the three novae are compared in Figure~\ref{fig_3novae},
a log-log plot between the time since discovery and the flux radiated
through the $V$ band normalized to the median value during the plateau.  To
compare the three novae on similar grounds, we selected to plot in
Figure~\ref{fig_3novae} the visual estimates collected by the AAVSO observers,
the naked eye being the only type of detector and photometric band in common
among the three novae.

The similarity of the three light-curves in Figure~\ref{fig_3novae} is
striking, especially the identical $\alpha$ slope through the decline
$F^{V}_{decline}~\propto~(t-t_0)^\alpha$.  All three novae:

	\begin{itemize}

	\item lingered initially for many months around maximum brightness,
	with persisting low-ionization conditions in their spectra,

	\item went through various re-brightening during the plateau,
	the first of them being invariably the brightest one,

	\item ended their plateau with a final brightening, the one showing
	the widest FWHM,

	\item experienced a rapid rise in the ionization degree upon
	embracing the steady decline from the plateau, with concomitant
	disappearance of the absorption lines and appearance of nebular
	[OIII] 5007, which was soon

	\item followed by appearance of [FeVII] and later by coronal [FeX].

	\end{itemize}

The [FeX] is not yet present in the spectra of NCas21 for day +660, but
given ($i$) the strict similarity to HR Del and V723 Cas, and ($ii$) the
monotonic increase in flux of [FeVII], we expect [FeX] to appear
at some later epoch in NCas21 too.

\section{Spectroscopy}

\subsection{Observations}

All the NCas21 spectra considered in this paper are high-resolution ones,
obtained with Echelle spectrographs.  A log-book of the spectroscopic
observations is presented in Table~\ref{tab_logbook}.  We collected a total
of 110 Echelle spectra distributed over 107 individual nights during the
time interval going from discovery to day +662 of the outburst of NCas21.

The vast majority of the Echelle spectra here considered have been obtained
with the Varese Schiaparelli Observatory 0.84m telescope, which is operated
by ANS Collaboration and it is equipped with an Astrolight Inst.  mk.III
Multi-Mode carbon-fiber spectrograph, that in the Echelle configuration
covers the 4250-8900 \AA\ range (see Munari \& Valisa (2014) for a
description of the optical train and a performance evaluation of early
models of these multi-mode spectrographs).  Cross-dispersion is achieved
with a N-SF11 prism, and the detector is a SBIG ST10XME CCD camera
(2192x1472 array, 6.8$\mu$m pixel, KAF-3200ME chip with micro-lenses to
boost the quantum efficiency).  The spectral resolution obtained with a R2
79 l/mm grating is 20,000 for a 1-arcsec slit and CCD binning=1$\times$,
lowering to 15,000 for a 2-arcsec slit, and to 12,000 for CCD
binning=2$\times$.  Only the two reddest orders are affected by limited
inter-order gaps (between 8239-8243 \AA\ and 8554-8574 \AA).

Additional spectra, particularly when NCas21 turned fainter at later
epochs, where collected with the 1.82m telescope + REOSC Echelle
spectrograph which is operated in Asiago by INAF (Italian National Institute
of Astrophysics).  The 3500-7350~\AA\ interval is covered on 32 orders
without inter-order gaps by an Andor DW436-BV camera (housing a 2048x2048
array, 13.5 $\mu$m pixel size, E2V CCD42-40 AIMO model CCD).  The resolving
power is 22,000 for the standard 1.8-arcsec slit-width.

A few further Echelle spectra were secured with the Stroncone 0.50m
telescope operated by ANS Collaboration, which is equipped with an
Astrolight Inst.  mk.III Multi-Mode spectrograph similar to that used in
Varese.  The detector is an ASI 1600MM CMOS camera (4656x3520 array,
3.8$\mu$m pixel, with a Panasonic MN34230 chip used in binning 2$\times$
mode).

All observations at all telescopes were conducted with the slit oriented
along the parallactic angle, for optimal sky-subtraction and
flux-calibration.  Data reduction for all telescopes has been performed in
IRAF and has included all usual steps for bias, dark, flat, sky subtraction,
wavelength calibration, heliocentric correction, and flux calibration.  The
latter has been achieved via nightly observations of spectrophotometric
standards, located on the sky close to NCas21.  This allowed to join all
Echelle orders into a single 1D-fluxed spectrum covering the whole recorded
wavelength range.  The zero-point of the each 1D-fluxed spectrum has been
checked against the $B$$V$$R$$I$ data for the same night as described in sect.~2
above.

	\begin{table}
	\centering
	\footnotesize
	\caption{Journal of the Echelle spectroscopic observations. The last column identify the
        contributing telescope: Varese 0.84m, Asiago 1.82m, and Stroncone 0.50m.}
	\begin{tabular}{ccrrcc|cccrrc}
        &&\\
	\hline
	&&&&&&&&&&&\\
	 date       & HJD          & t-t$_{\rm 0}$& expt & tel.  &&  date       & HJD          & t-t$_{\rm 0}$& expt & tel. \\
	            &              &              &(sec) &       &&             &              &              &(sec) &     \\
	&&&&&&&&&&&\\
	\hline
	&&&&&&&&&&&\\
 2021-03-19 & 2459293.29 &   0.37 & 1800 & 0.84m && 2021-10-16 & 2459504.24 & 211.32 & 3150 & 0.84m \\
 2021-03-20 & 2459294.28 &   1.35 & 2700 & 0.84m && 2021-10-24 & 2459512.25 & 219.33 & 4500 & 0.84m \\
 2021-03-21 & 2459295.28 &   2.35 & 2700 & 0.84m && 2021-11-04 & 2459523.38 & 230.46 & 3600 & 0.84m \\
 2021-03-22 & 2459296.29 &   3.37 & 2700 & 0.84m && 2021-11-12 & 2459351.34 & 238.42 & 2400 & 0.84m \\
 2021-03-23 & 2459297.41 &   4.49 & 1800 & 0.84m && 2021-11-18 & 2459537.28 & 244.35 & 3600 & 0.84m \\
 2021-03-24 & 2459298.29 &   5.37 & 2700 & 0.84m && 2021-11-29 & 2459548.35 & 255.43 & 3600 & 0.84m \\
 2021-03-24 & 2459298.26 &   5.34 &  900 & 1.82m && 2021-12-03 & 2459552.21 & 259.29 & 3600 & 0.84m \\
 2021-03-25 & 2459299.33 &   6.41 & 3600 & 0.84m && 2021-12-06 & 2459555.21 & 262.29 & 4050 & 0.84m \\
 2021-03-27 & 2459301.31 &   8.38 & 4800 & 0.84m && 2021-12-07 & 2459556.22 & 263.29 & 4800 & 0.84m \\
 2021-03-28 & 2459302.31 &   9.38 & 2700 & 0.84m && 2021-12-09 & 2459558.27 & 265.35 & 4800 & 0.84m \\
 2021-03-29 & 2459303.29 &  10.37 & 1800 & 0.84m && 2021-12-11 & 2459560.14 & 267.32 & 6300 & 0.84m \\
 2021-03-30 & 2459304.29 &  11.36 & 2700 & 0.84m && 2021-12-13 & 2459562.39 & 269.46 & 3600 & 0.84m \\
 2021-03-31 & 2459305.29 &  12.36 & 2700 & 0.84m && 2021-12-15 & 2459564.29 & 271.37 & 3600 & 0.84m \\
 2021-04-01 & 2459306.29 &  13.36 & 2700 & 0.84m && 2021-12-17 & 2459566.26 & 273.34 & 3600 & 0.84m \\
 2021-04-03 & 2459308.29 &  15.37 & 2700 & 0.84m && 2021-12-18 & 2459567.25 & 274.33 &  600 & 1.82m \\ 
 2021-04-04 & 2459309.29 &  16.37 & 2700 & 0.84m && 2021-12-20 & 2459569.22 & 276.30 & 5400 & 0.84m \\
 2021-04-05 & 2459310.29 &  17.37 & 2700 & 0.84m && 2021-12-22 & 2459571.29 & 278.33 &  600 & 1.82m \\
 2021-04-07 & 2459312.29 &  19.37 & 2700 & 0.84m && 2021-12-22 & 2459571.35 & 278.39 & 5400 & 0.84m \\
 2021-04-08 & 2459313.29 &  20.37 & 2700 & 0.84m && 2021-12-26 & 2459575.31 & 282.39 & 5400 & 0.84m \\
 2021-04-13 & 2459318.31 &  25.39 & 1800 & 0.84m && 2021-12-28 & 2459577.35 & 284.43 & 6300 & 0.84m \\
 2021-04-17 & 2459321.66 &  26.74 &  900 & 0.84m && 2021-12-30 & 2459579.22 & 286.29 & 4500 & 0.84m \\
 2021-04-17 & 2459322.30 &  29.38 & 2400 & 0.84m && 2021-12-31 & 2459580.23 & 287.31 & 4500 & 0.84m \\
 2021-04-18 & 2459323.30 &  30.38 & 2400 & 0.84m && 2022-01-06 & 2459586.24 & 293.31 & 5400 & 0.84m \\
 2021-04-19 & 2459324.31 &  31.39 & 1500 & 0.84m && 2022-01-09 & 2459589.25 & 296.33 & 5400 & 0.84m \\
 2021-04-21 & 2459326.32 &  33.40 & 2250 & 0.84m && 2022-01-11 & 2459591.26 & 298.33 & 5400 & 0.84m \\
 2021-04-22 & 2459327.31 &  34.38 & 2250 & 0.84m && 2022-01-14 & 2459594.26 & 301.33 & 5400 & 0.84m \\
 2021-04-23 & 2459327.64 &  34.72 & 2250 & 0.84m && 2022-01-20 & 2459600.25 & 307.33 & 5400 & 0.84m \\
 2021-04-24 & 2459328.31 &  35.38 & 2400 & 0.84m && 2022-01-21 & 2459601.22 & 308.30 &  600 & 1.82m \\
 2021-05-03 & 2459337.62 &  44.69 & 1500 & 0.84m && 2022-01-27 & 2459607.26 & 314.34 & 5400 & 0.84m \\
 2021-05-08 & 2459342.58 &  49.66 & 1200 & 0.84m && 2022-02-01 & 2459612.26 & 319.34 & 5400 & 0.84m \\
 2021-05-12 & 2459346.64 &  53.72 & 1380 & 0.84m && 2022-02-08 & 2459619.29 & 326.36 & 5400 & 0.84m \\
 2021-05-17 & 2459351.62 &  58.70 &  900 & 0.84m && 2022-02-11 & 2459621.29 & 328.36 &  600 & 1.82m \\
 2021-05-19 & 2459354.39 &  61.47 & 1500 & 0.84m && 2022-02-21 & 2459632.28 & 339.36 & 5400 & 0.84m \\
 2021-05-25 & 2459360.36 &  67.43 & 2100 & 0.84m && 2022-02-23 & 2459634.28 & 341.36 & 5400 & 0.84m \\
 2021-05-27 & 2459362.35 &  69.42 & 2400 & 0.84m && 2022-02-28 & 2459639.28 & 346.36 & 5400 & 0.84m \\
 2021-05-28 & 2459363.36 &  70.43 & 1500 & 0.84m && 2022-03-06 & 2459645.28 & 352.36 & 5400 & 0.84m \\
 2021-05-31 & 2459366.38 &  73.46 & 1500 & 0.84m && 2022-03-20 & 2459659.30 & 366.38 & 5400 & 0.84m \\
 2021-06-04 & 2459370.36 &  77.44 & 2100 & 0.84m && 2022-03-29 & 2459668.32 & 375.39 & 6300 & 0.84m \\
 2021-06-08 & 2459374.37 &  81.45 &  900 & 0.84m && 2022-04-05 & 2459675.32 & 382.40 & 5400 & 0.84m \\
 2021-06-10 & 2459376.39 &  83.47 &  900 & 0.84m && 2022-04-10 & 2459680.32 & 387.40 & 4500 & 0.84m \\
 2021-06-13 & 2459379.38 &  86.46 &  630 & 0.84m && 2022-04-18 & 2459688.33 & 395.41 & 5400 & 0.84m \\
 2021-06-17 & 2459383.38 &  90.46 & 1200 & 0.84m && 2022-05-11 & 2459711.38 & 418.44 & 2700 & 0.84m \\
 2021-06-24 & 2459390.44 &  97.51 & 1200 & 0.84m && 2022-05-17 & 2459717.42 & 424.50 & 5400 & 0.84m \\
 2021-06-28 & 2459394.35 & 101.43 & 1260 & 0.50m && 2022-06-25 & 2459756.40 & 463.48 & 4500 & 0.84m \\
 2021-07-05 & 2459401.38 & 108.46 &  840 & 0.50m && 2022-07-10 & 2459771.40 & 478.48 & 5400 & 0.84m \\
 2021-07-09 & 2459405.42 & 113.50 &  900 & 0.50m && 2022-08-08 & 2459800.36 & 507.43 & 5400 & 0.84m \\	
 2021-07-18 & 2459414.38 & 122.46 &  360 & 1.82m && 2022-09-25 & 2459848.34 & 555.42 & 7200 & 0.84m \\
 2021-08-08 & 2459435.33 & 142.40 & 1200 & 0.84m && 2022-10-10 & 2459863.39 & 570.47 &  900 & 1.82m \\	
 2021-08-17 & 2459444.35 & 151.43 &  900 & 0.84m && 2022-10-28 & 2459866.35 & 588.43 & 6300 & 0.84m \\		
 2021-08-19 & 2459446.39 & 153.47 &  180 & 1.82m && 2022-11-11 & 2459895.37 & 602.45 &  900 & 1.82m \\	 
 2021-08-23 & 2459450.35 & 157.43 & 2700 & 0.84m && 2022-11-23 & 2459907.37 & 614.45 & 7200 & 0.84m \\
 2021-09-01 & 2459459.35 & 166.42 & 3600 & 0.84m && 2022-12-05 & 2459919.33 & 626.41 & 1200 & 1.82m \\
 2021-09-11 & 2459469.37 & 176.45 & 1800 & 0.84m && 2022-12-21 & 2459935.35 & 642.43 & 6300 & 0.84m \\
 2021-09-30 & 2459488.27 & 195.34 & 3150 & 0.84m && 2023-01-05 & 2459950.39 & 657.47 & 7200 & 0.84m \\
 2021-10-06 & 2459494.33 & 201.40 & 3600 & 0.84m && 2023-01-10 & 2459955.40 & 662.50 & 5400 & 1.82m \\
	&&&&&&&&&&&\\
	\hline
	\end{tabular}
	\label{tab_logbook}
	\end{table}

While the integrated flux of emission lines and of the in-between continuum
has been measured on the 1D-fluxed spectra, a continuum-normalized version
of the same 1D-spectra has been used instead to prepare the pictures that in
this papers illustrate the overall spectral evolution.  The emission lines
of NCas21 are so wide to cover a significant fraction of a single Echelle
order or are split among two adjacent ones, and our effort to
continuum-normalize the spectra has been made much easier and far more
objective by the availability of their 1D-fluxed version.

\subsection{Dynamic-plots of the spectral evolution}

\begin{figure}[!ht]
	\centering
	\includegraphics[width=15.8cm]{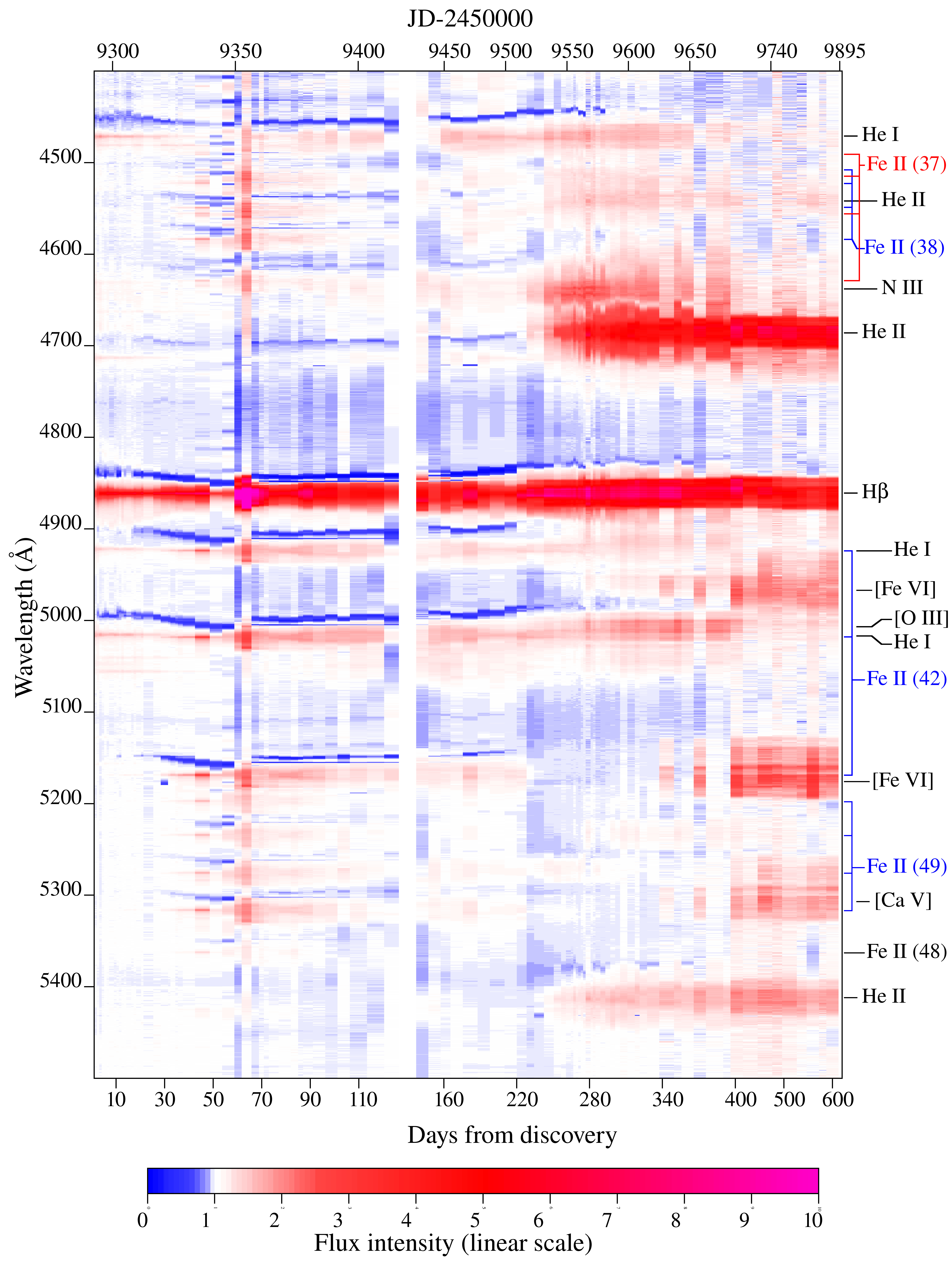}
	\caption{Sequence of Echelle spectra for the spectral range 4400 $-$
	5500 \AA.  The spectra are continuum normalized, with the white
	color corresponding to 1.0.  The observing epochs are marked as days
	passed since discovery (cf Eq.  (1)).  The temporal cadence is 1-day
	for the first 130 days, 2-days for day +130 to +400, and 5-days
	afterward, in pace with the slowing evolution of the nova and the
	lower S/N of the spectra at late epochs that would benefit from
	averaging them together.}
	\label{fig_dynamic_a}
	\end{figure}

\begin{figure}[!ht]
	\centering
	\includegraphics[width=16.8cm]{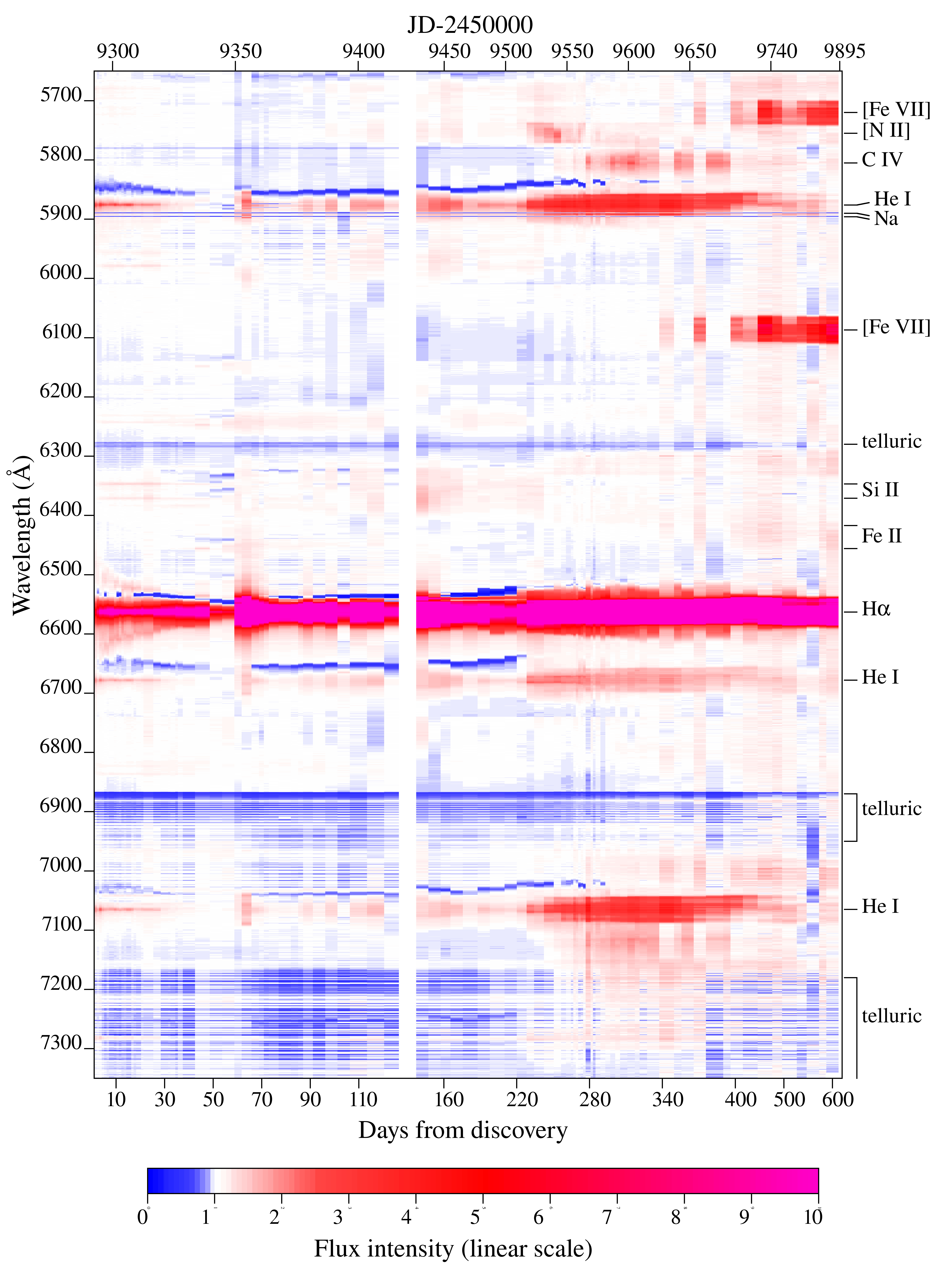}
	\caption{Similarly to Figure~\ref{fig_dynamic_a} for the spectral
	range 5650 $-$ 7350 \AA.}
	\label{fig_dynamic_b}
	\end{figure}

\begin{figure}[!ht]
	\centering
	\includegraphics[width=16.3cm]{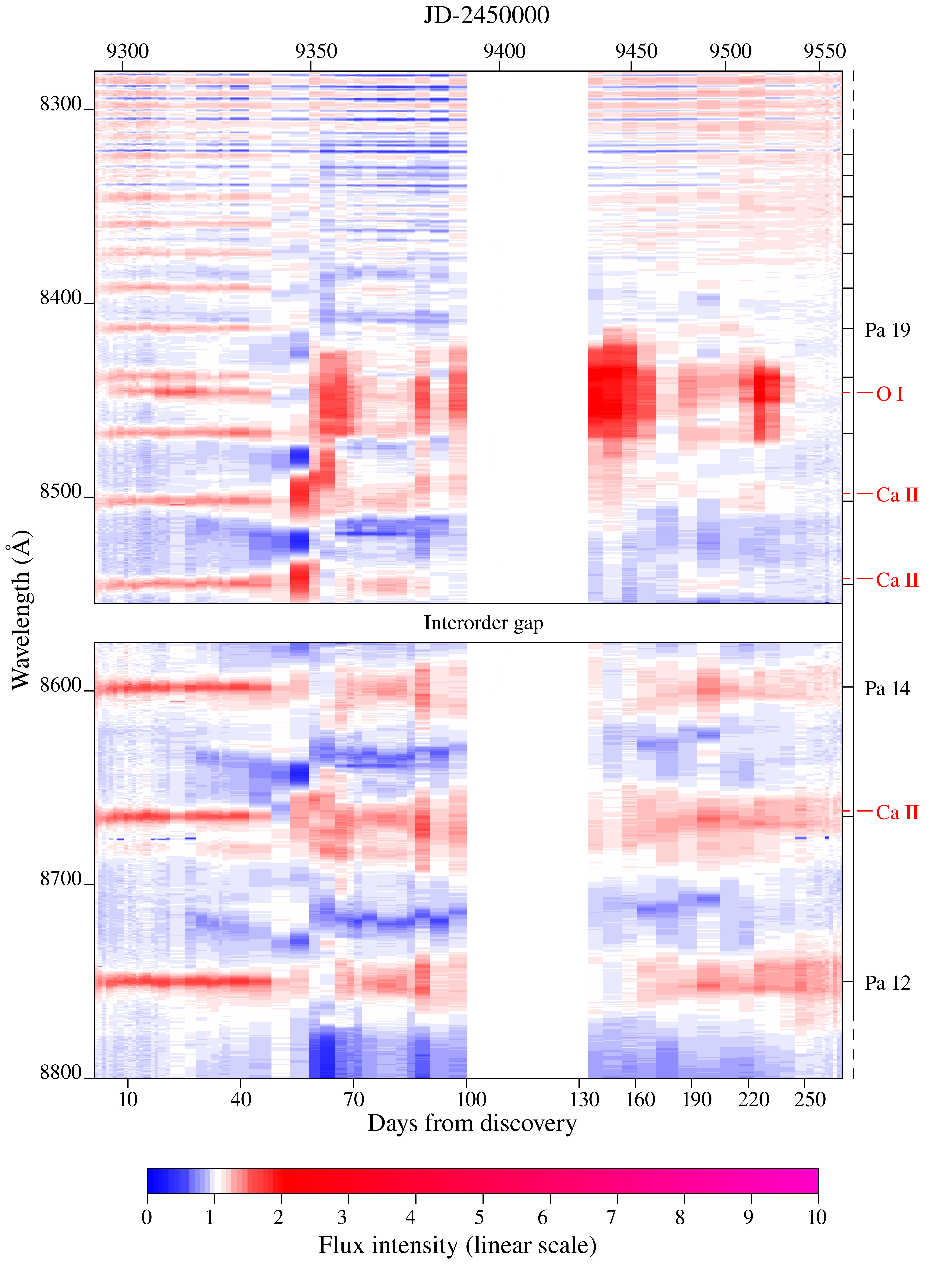}
	\caption{Similarly to Figure~\ref{fig_dynamic_a} for the spectral
        range 8280 $-$ 8800 \AA\ (with an inter-order gap between 8555 and
        8575 \AA).  The time interval shown is limited to the first 270 days
        of the nova evolution because of the poorer S/N at such far-red
        wavelengths for the spectra recorded at later epochs.}
	\label{fig_dynamic_c}
	\end{figure}

\begin{figure}[!ht]
	\centering
	\includegraphics[width=16.3cm]{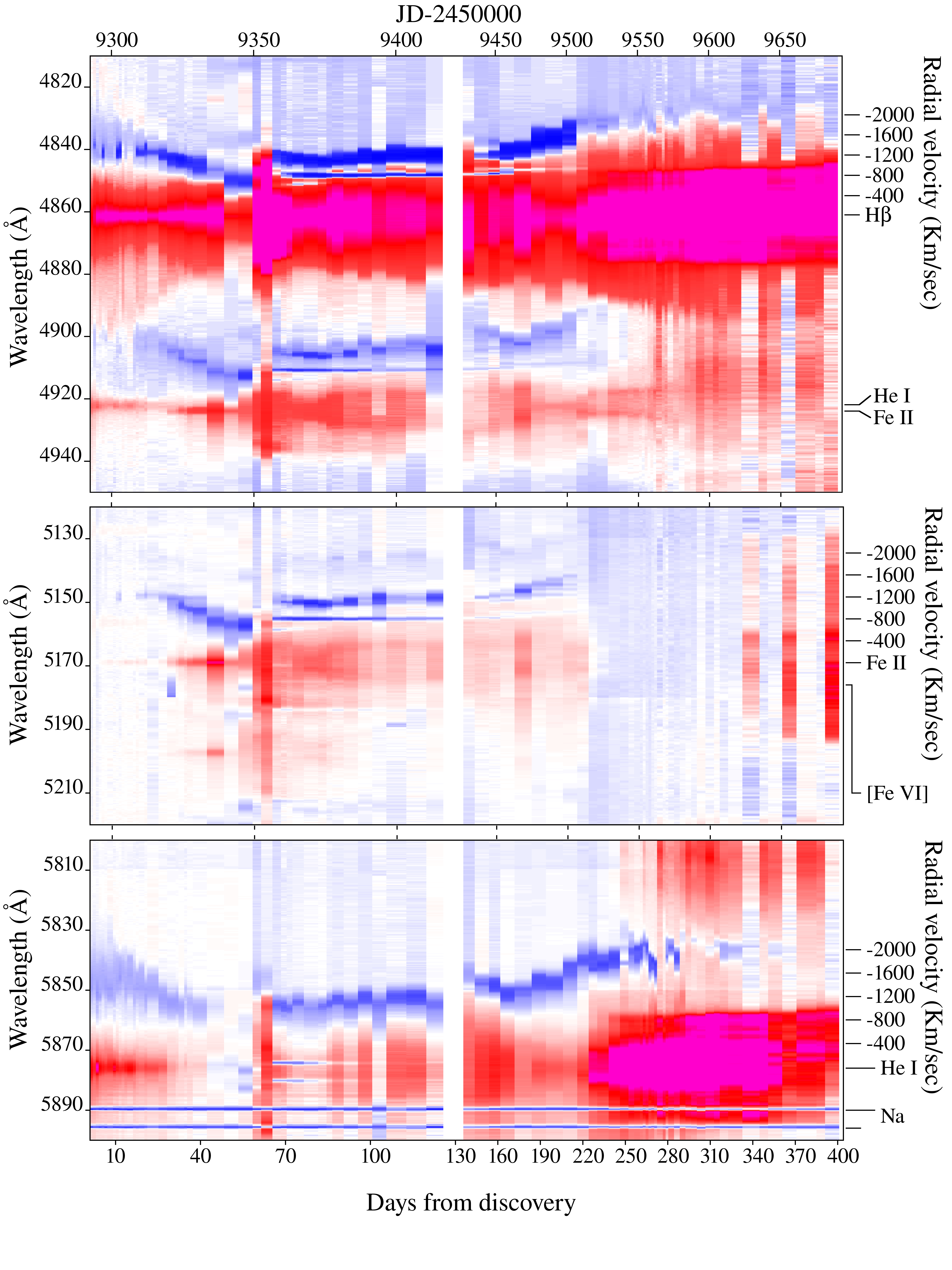}
	\caption{Enlarged portions from Figures~\ref{fig_dynamic_a} and 
        \ref{fig_dynamic_b} to 	highlight the behavior of P-Cyg absorptions.  
        The time interval does 	not extend beyond day +400 because all P-Cyg 
        absorptions vanished by day +320.  Top panel: H$\beta\ $ + FeII(42) 
        4924 blended with HeI 4922 .  
        Middle panel: FeII 5169 (mult.  42), with flaring of [FeVI]
	after day +350.  Bottom panel: HeI 5876, with both stellar and
	interstellar NaI doublet.}
	\label{fig_dynamic_d}
	\end{figure}

Dynamic-plots offer a practical way to present at a glance the information
stored in a long series of spectra which are part of a temporal sequence. 
We have used this technique in
Figures~\ref{fig_dynamic_a}$-$~\ref{fig_dynamic_c} to illustrate,
respectively, the evolution of NCas21 over the HeII/H$\beta$/FeII-42,
[FeVII]/H$\alpha$/HeI, and OI/Paschen-head/CaII-triplet regions, and in
Figure~\ref{fig_dynamic_d} to zoom in the behavior of P-Cyg absorptions to
H$\beta$/FeII/HeI.

Such plots are prepared from the continuum-normalized version of the 1D-fluxed
spectra, where the flat continuum is colored white, the absorptions in shades of 
blue and the emission in shades of pink and red, as illustrated by the color-bar
at the bottom of each figure.

The time-scale of the dynamic-plots in
Figures~\ref{fig_dynamic_a}$-$~\ref{fig_dynamic_d} is itself dynamic.  The
temporal cadence is 1-day for the first 130 days, to both reflect the daily
pace of our observations of NCas21 as well as to better map the rapid
changes exhibited by the spectral features during the initial plateau.  The
cadence turns to 2-days for the time-interval from day +130 to +400, and
slows to 5-days for day +400 to day +662, following the slower nova
evolution and the lower S/N of the spectra that benefit from averaging
them together over 5-day bins.

\subsection{Measuring the reddening from the spectra}

The spectra of NCas21 present strong interstellar features which allow to
estimate the amount of reddening affecting the nova.  We carried out already
this exercise (Munari et al.  2021a) on the first Echelle
spectrum we secured within hours of the nova announcement (day +0.37),
deriving E(B-V)=0.53.  Here we refine the reddening estimates by repeating
the measurements on the best available spectra and averaging the results,
which indeed do little differ from the initial ones.

The atomic interstellar lines presents two distinct components, at
heliocentric velocities $-$55.3 $\pm$0.7 and $-$14.4 $\pm$0.8 km/s.  For the
NaI D1 line, their equivalent widths are respectively 0.173 and 0.740 \AA,
with the latter appearing clearly saturated.  Adopting the Munari \& Zwitter
(1997) calibration, the reddening from the unsaturated NaI D1 component at
$-$55.3 km/s is E(B-V)=0.055, and E(B-V)=0.45 is obtained from the
unsaturated KI component at -14.4 km/s showing an equivalent width of
0.118~\AA, for a total reddening of E(B-V)=0.51.  The diffuse interstellar
band at 6614.5~\AA\ is characterized by an equivalent width of 0.112~\AA,
which translates to E(B-V)=0.49 following the calibration by Munari (2014). 
By averaging them, we adopt in this paper 
\begin{equation} 
E(B-V) = 0.50 \pm 0.01 
\end{equation} 
as the interstellar reddening affecting NCas21.

This value of the reddening compares well with the $(B-V)_{\rm
max}$=+0.662 $\pm$0.010 (Eq.4 above) displayed by NCas21 at maximum.  Van
den Bergh \& Younger (1987) have compiled and analyzed UBV photometry of
classical novae in eruption, deriving an intrinsic $(B-V)_\circ$=0.23
$\pm$0.06 at maximum brightness, which translates into E(B-V)=0.43
$\pm$0.06 for NCas21.

\subsection{Evolution during pre-maximum}

NCas21 has presented a smooth evolution during the flat phase leading up to
the maximum on day +53.  The emission spectrum, up to day +20 from
discovery, was dominated by Balmer lines and weakening of HeI lines, with a
velocity around $-$1500 km/s for the single-component P-Cyg absorptions. 
The fading of HeI lines was accompanied by the appearance of FeII emission
lines, with the earliest detection for multiplet 42 on day +8.4, while the
first appearance of the corresponding FeII P-Cyg absorptions occurring much
later on day +26.7\footnote{\it all dates hereafter refer to the corresponding
spectrum in Table~\ref{tab_logbook}.}.  During this phase the FeII emission
lines remained weak and very narrow (Figure~\ref{fig_dynamic_d}, central
panel) with FWHM$\approx$160~km/s, concomitant with the presence of HeI
emission lines declining in intensity.  The FeII P-Cyg absorptions were
centered at $-$1180 km/s and characterized by FWHM$\approx$190~km/s, well
detached from the corresponding emission component positioned close to the
laboratory wavelength.  From day +31.4 until the passage at maximum
brightness on day +53, the FeII emission experienced a 5-fold increase, with
all the usual multiplets 27, 29, 37, 38, 41, 42, 46, 49, 55, 73, 74 present
and complete.

As illustrated by Figure~\ref{fig_dynamic_d} and Figure~\ref{fig_P_Cyg}, the
ejecta maintained turbulent fluctuations in velocity and density on a daily
pace up to day +16.4, when the overall velocity of the P-Cyg absorption of
all lines (cf.  Figure~\ref{fig_dynamic_a}) begun a smooth decline from
$-$1500 km/s to $-$700 km/s, which was reached at the time of maximum brightness
(day +53).  In parallel, also the edge-velocity kept declining, from $-$2500
km/s on day +5.4, to $-$1700, $-$1300, and $-$1000 km/sec on day
16.4, +35.4, and +49.7, respectively.

The profile of a few representative emission lines for days +5.4 and +29.5
are presented in the first two panels of Figure~\ref{fig_12_date}, with the
shape of the P-Cyg absorption suggesting the presence of a large velocity
gradient in a diluted medium.  It is also worth noticing the different line
profile between HeI triplet and singlet lines, because of the collisional
excitation of meta-stable 2S$^{3}$ state in the dense ejecta environment, with
much larger wings for triplet lines: on day +5.4 the FWHM of
triplet HeI 5876 is 630 km/s, almost three times that of singlet HeI 5016 that
reaches only 230 km/s.  The evolution of the reddening-corrected 5876/6078
flux ratio for NCas21 is plotted in the bottom panel of
Figure~\ref{fig_dereddened}.

The evolution of OI 8446 line is presented in Figure~\ref{fig_dynamic_c}:
the line, located next to Paschen 18, is missing on day +0.37, appears on
day +1.35, and then remains rather stable through the whole pre-maximum.  The
evolution of the reddening-corrected OI 8446/7774 flux ratio is presented on
the top panel of Figure~\ref{fig_dereddened}, and it is clearly
characterized by two distinct phases: during pre-maximum, the ratio evolved
slowly and smoothly, remaining between 0.7 and 2.1, but soon after maximum
brightness, the ratio started to jump up-and-down widely, up to
8446/7774$\sim$25 around day +150, a value indicating a dominant role played
by Lyman-$\beta$ fluorescence in pumping the 8446 line (Bhatia \& Kastner
(1995)).  Nebular [OI] 6300, 6364 and auroral 5577 lines have never turned
visible on NCas21 spectra during the monitoring period.

\subsection{Passage at primary maximum}

From the photometric point of view, and similarly to HR Del and V723 Cas in
Figure~\ref{fig_3novae}, also in the case of NCas21 the rise to, passage at,
and decline from maximum appears as an isolated and quickly-evolving event,
superimposed on a protracted plateau.  For NCas21, the rise begun on day
+45, with maximum was reached a week later on day +53, and the decline
commencing immediately and the nova returning to the flat plateau brightness
by day +62.

	\begin{figure} 
	\centering
	\includegraphics[width=13.6cm]{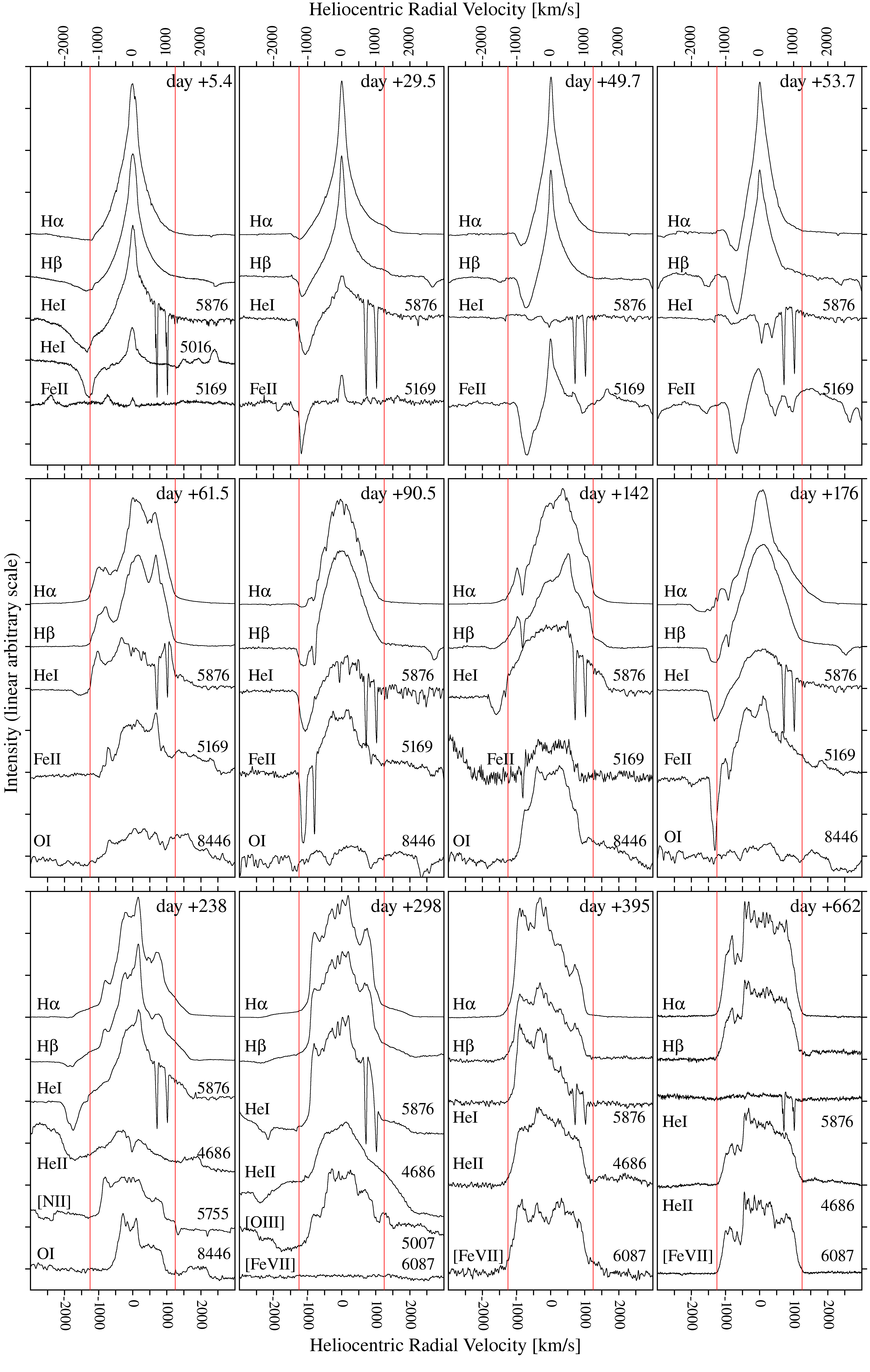}
	\caption{Representative steps along the spectral evolution and
	Doppler broadening of some the most relevant emission lines observed
	in Nova Cas 2021.  The spectra are plotted on a linear scale, with
	an arbitrary offset to avoid overlap.  The vertical red lines mark
	$\pm$1250~km/s velocities.  \textit{Top row:} H$\alpha$, H$\beta$,
	HeI (triplet and singlet), FeII 5169 and OI 8446 from discovery to
	maximum brightness (day +53).  \textit{Middle row:} H$\alpha$,
	H$\beta$, HeI, FeII 5169 and OI 8446 during the plateau phase and
	secondary maxima.  Day +176 corresponds to a secondary maximum,
	while day +142 is representative of the underlying plateau away from
	the secondary maxima.  \textit{Bottom row:} H$\alpha$, H$\beta$, HeI,
	[NII], [OIII], [FeVII] and OI~8446 during the final decline after
	the nova abandoned the plateau on day +230.}
	\label{fig_12_date} 
	\end{figure}

	\begin{table}
	\caption{Logarithm of the flux in the continuum at 4770~\AA\ (in
        erg~cm$^{-2}$ s$^{-1}$ \AA$^{-1}$) and the logarithm of the
        integrated flux (in erg~ cm$^{-2}$ s$^{-1}$) of HeII 4686 ,
        H$\beta$ , [OIII] 5007 , FeII 5169 , [NII] 5755 , HeI 5876
        (triplet), HeI 6678 (singlet), [FeVII] 6087 ,
        H$\alpha$, and OI 8446 emission lines from the Echelle spectra
        listed in Table~\ref{tab_logbook}.} 
	\centering
	\footnotesize
	\begin{tabular}{lrrrrrrrrrrr}
        &&\\
	\hline
	&&\\
	\multicolumn{1}{c}{t-t$_{\rm 0}$}&
	\multicolumn{1}{c}{HeII~~}&
	\multicolumn{1}{c}{4770 \AA}&
	\multicolumn{1}{c}{H$\beta$~~}&
	\multicolumn{1}{c}{[O III]~~}&
	\multicolumn{1}{c}{Fe II}&
	\multicolumn{1}{c}{[NII]}&
	\multicolumn{1}{c}{HeI 5876 }&
	\multicolumn{1}{c}{[Fe VII]}&
	\multicolumn{1}{c}{H$\alpha$}&
	\multicolumn{1}{c}{HeI 6678}&
	\multicolumn{1}{c}{OI 8846 } \\
	&&\\
	\hline
	&&\\

0.37 & 		& -11.59 & -9.96 & 	&  	  & 	&-10.18 & & -9.49 & -10.57 & 	\\
1.35 & 		& -11.40 & -9.75 & 	&  	  & 	&-10.17 & & -9.26 & -10.61 & -11.53 \\
2.35 & 		& -11.48 & -9.79 & 	&	  &     &-10.18 & & -9.22 & -10.63 & -11.82 \\
3.37 & 		& -11.47 & -9.79 & 	&	  &     &-10.28 & & -9.21 & -10.79 & -11.66 \\
4.49 & 		& -11.45 & -9.77 & 	&  -12.55 &     &-10.35 & & -9.18 & -10.87 & -11.55 \\
5.37 & 		& -11.48 & -9.74 & 	&  -12.33 &     &-10.31 & & -9.12 & -10.86 & -11.54 \\
6.41 & 		& -11.56 & -9.86 & 	&  -12.25 &     &-10.36 & & -9.15 & -10.86 & -11.49 \\
8.38 & 		& -11.61 & -9.94 & 	&  -12.73 &     &-10.48 & & -9.23 & -10.84 & -11.53 \\
9.38 & 		& -11.55 & -9.89 & 	&  -12.50 &     &-10.46 & & -9.24 & -10.94 & -11.43 \\
10.37 & 	& -11.52 & -9.94 & 	&  -12.24 &     &-10.54 & & -9.31 & -10.99 & -11.43 \\
11.36 & 	& -11.56 & -9.93 & 	&  -12.47 &     &-10.51 & & -9.26 & -10.97 & -11.59 \\
12.36 & 	& -11.55 & -9.93 & 	&  -12.18 &     &-10.46 & & -9.29 & -10.96 & -11.40 \\
13.36 & 	& -11.57 & -9.99 & 	&  -12.28 &     &-10.52 & & -9.33 & -10.96 & -11.42 \\
15.37 & 	& -11.57 & -9.97 & 	&  -12.40 &     &-10.45 & & -9.29 & -10.89 & -11.37 \\
16.37 & 	& -11.61 & -9.93 & 	&  -12.19 &     &-10.50 & & -9.28 & -11.00 & -11.29 \\
17.37 & 	& -11.56 & -9.94 & 	&  -12.38 &     &-10.53 & & -9.28 & -10.97 & -11.27 \\
19.37 & 	& -11.60 & -9.97 & 	&  -12.24 &     &-10.56 & & -9.27 & -11.00 & -11.21 \\
20.37 & 	& -11.57 & -10.05 & 	&  -12.52 &     &-10.70 & & -9.36 & -11.05 & -11.28 \\
25.39 & 	& -11.52 & -9.87 & 	&  -12.23 &     &-10.56 & & -9.19 & -10.95 & -11.20 \\
26.74 & 	& -11.59 & -9.93 & 	&  -12.10 &     &-10.74 & & -9.16 & -11.18 & -11.39 \\
29.38 & 	& -11.53 & -9.78 & 	&  -11.62 &     &-10.62 & & -9.11 & -11.03 & -11.31 \\
30.38 & 	& -11.48 & -9.74 & 	&  -11.92 &     &-10.81 & & -9.10 & -11.16 & -11.33 \\
31.39 & 	& -11.46 & -9.72 & 	&  -11.10 &     &-10.86 & & -9.08 & -11.19 & -11.48 \\
33.40 & 	& -11.52 & -9.74 & 	&  -11.12 &     &-10.86 & & -9.00 & -11.24 & -11.23 \\
34.72 & 	& -11.58 & -9.77 & 	&  -10.93 &     &-10.78 & & -9.01 & -11.26 & -11.27 \\
35.38 & 	& -11.50 & -9.72 & 	&  -10.85 &     &-10.95 & & -9.02 & -11.41 & -11.01 \\
44.69 & 	& -11.30 & -9.44 & 	&  -10.23 &     &-11.22 & & -8.82 &  	   & -11.00 \\
49.66 & 	& -10.73 & -9.51 & 	&  -10.09 & 	&       & & -8.81 & 	   & -10.72 \\
53.72 & 	& -9.51 & -9.27 & 	&  -9.95 & 	&       & & -8.81 & 	   & -10.44  \\
58.70 &		& -11.55 & -8.96 & 	&  -10.09 &     &-10.14 & & -8.19 & -10.46 & -9.90 \\
61.47 & 	& -11.70 & -8.99 &	&  -10.09 &     &-9.83  & & -8.17 & -10.27 & -9.74 \\
67.43 & 	& -11.58 & -9.22 & 	&  -10.33 &     &-10.25 & & -8.26 & -10.79 & -10.10 \\
69.42 & 	& -11.45 & -9.28 & 	&  -10.33 &     &-10.24 & & -8.34 & -10.68 & -10.27 \\
70.43 & 	& -11.55 & -9.46 & 	&  -10.37 &     &-10.40 & & -8.37 & -10.94 & -10.42 \\
73.46 & 	& -11.44 & -9.50 & 	&  -10.19 &     &-10.41 & & -8.48 & -10.96 & -10.70 \\
77.44 & 	& -11.27 & -9.38 & 	&  -10.16 &     &-10.27 & & -8.49 & -10.76 & -10.94 \\
81.45 & 	& -11.26 & -9.30 & 	&  -10.15 &     &-10.32 & & -8.44 & -10.77 & -11.17 \\
83.47 & 	& -11.39 & -9.25 & 	&  -10.20 &     &-10.07 & & -8.34 & -10.52 & -10.60 \\
86.46 & 	& -11.51 & -9.31 & 	&  -10.35 &     &-10.15 & & -8.33 & -10.51 & -10.15 \\
90.46 & 	& -11.38 & -9.36 & 	&  -10.20 &     &-10.16 & & -8.39 & -10.58 & -10.80 \\
97.51 & 	& -11.52 & -9.50 & 	&  -10.54 & -10.90    &-10.12 & & -8.45 & -10.48 & -10.22 \\
101.4 & 	& -11.14 & -9.33 & 	&  -10.44 &     &-10.24 & & -8.62 & -10.51 & 	\\
108.5 & 	& -11.60 & -9.63 & 	&  -10.65 & -11.04    &-10.25 & & -8.58 & -10.50 & 	\\
113.5 & 	& -11.61 & -9.65 & 	&  -10.85 & -11.07    &-10.29 & & -8.73 & -10.49 & 	\\
122.5 & 	& -11.22 & -9.53 & 	&  -10.59 &     &-10.27 & & -8.76 & -10.56 & 	\\
142.4 & -11.24	& -11.82 & -9.60 & 	&  -10.68 & -10.82    &-10.16 & & -8.42 & -10.56 & -10.07 \\
151.4 & 	& -11.71 & -9.79 & 	&  -10.66 & -10.90    &-10.10 & & -8.61 & -10.36 & -10.10 \\
157.4 & 	& -11.66 & -9.60 & 	&  -10.74 & -10.85    &-10.01 & & -8.76 & -10.34 & -10.09 \\
166.4 & 	& -11.74 & -9.84 & 	&  -11.03 & -10.87    &-10.24 & & -9.00 & -10.57 & -10.48 \\
176.5 & 	& -11.32 & -9.16 & 	&  -10.10 &     &-9.97  & & -8.50 & -10.45 & -10.69 \\
195.3 & 	& -11.53 & -9.63 & 	&  -10.86 &     &-10.22 & & -8.70 & -10.55 & -10.59 \\
201.4 & 	& -11.57 & -9.71 & 	&  -10.74 &     &-10.30 & & -8.94 & -10.77 & -10.87 \\
211.3 & 	& -11.36 & -9.50 & 	&  -10.41 &     &-10.21 & & -8.79 & -10.59 & -10.80 \\
219.3 & 	& -11.45 & -9.37 & 	&  -10.45 & -11.15 &-10.02 & & -8.56 & -10.46 & -10.43 \\
230.5 & -10.66 & -11.84 & -9.63 & 	&  -10.95 & -10.75 & -10.16 & & -8.89 & -10.49 & -10.55 \\
238.4 & -10.40 & -11.80 & -9.63 & 	&  -11.10 & -10.52 & -10.14 & & -8.90 & -10.41 & -10.81 \\
244.3 & -10.30 & -11.93 & -9.70 & 	& 	  & -10.59 & -10.14 & & -9.00 & -10.52 & -11.61 \\
255.4 & -10.35 & -12.11 & -9.89 & 	& 	  & -10.84 & -10.34 & & -9.22 & -10.72 &  \\
255.4 & -10.21 & -12.11 & -9.86 & 	& 	  & -11.17 & -10.26 & & -9.21 & -10.70 &  \\
259.3 & -10.24 & -12.11 & -9.89 & 	& 	  & -11.10 & -10.29 & & -9.24 & -10.71 &  \\
263.3 & -10.07 & -12.14 & -9.88 & 	& 	  & -11.23 & -10.29 & & -9.25 & -10.82 &  \\
267.3 & -10.11 & -12.05 & -9.80 & 	& 	  & -11.07 & -10.31 & & -9.27 & -10.66 &  \\
&&\\
	\hline
	\end{tabular}
	\label{tab_log}
	\end{table}
\clearpage
\setcounter{table}{3}

	\begin{table}
	\caption{(continued).}
	\centering
	\footnotesize
	
	\begin{tabular}{lrrrrrrrrrrr}
	
        &&\\
	\hline
	&&\\
	\multicolumn{1}{c}{t-t$_{\rm 0}$}&
	\multicolumn{1}{c}{HeII~~}&
	\multicolumn{1}{c}{4770 \AA}&
	\multicolumn{1}{c}{H$\beta$~~}&
	\multicolumn{1}{c}{[O III]~~}&
	\multicolumn{1}{c}{Fe II}&
	\multicolumn{1}{c}{[NII]}&
	\multicolumn{1}{c}{HeI 5876 }&
	\multicolumn{1}{c}{[Fe VII]}&
	\multicolumn{1}{c}{H$\alpha$}&
	\multicolumn{1}{c}{HeI 6678 }&
	\multicolumn{1}{c}{OI 8846} \\
	&&\\
	\hline
	&&\\
273.3 & -10.24 & -12.18 & -9.94 & 	& 	  & -11.38 & -10.35 & & -9.34 & -10.70 &  \\
276.3 & -10.30 & -12.40 & -10.12 & -11.09 &       & 	& -10.54 &        & -9.46 & -11.05 &  \\
278.4 & -10.20 & -12.27 & -10.05 & -11.01 &       & 	& -10.46 &        & -9.45 & -10.98 &  \\
282.4 & -10.08 & -12.20 & -9.98  & -10.92 &       & 	& -10.44 &        & -9.40 & -10.97 &  \\
286.3 & -10.18 & -12.33 & -10.07 & -11.02 &       & 	& -10.51 &        & -9.45 & -10.90 &  \\
293.3 & -10.18 & -12.32 & -10.09 & -11.06 &       & 	& -10.52 &        & -9.47 & -11.00 &  \\
296.3 & -10.13 & -12.37 & -10.10 & -11.06 &       & 	& -10.57 &        & -9.52 & -11.09 &  \\
298.3 & -10.14 & -12.40 & -10.11 & -11.00 &       & 	& -10.56 &        & -9.51 & -11.06 &  \\
301.3 & -10.21 & -12.39 & -10.13 & -10.99 &       & 	& -10.58 &        & -9.52 & -11.11 &  \\
307.3 & -10.24 & -12.44 & -10.16 & -11.05 &       & 	& -10.58 &        & -9.51 & -11.10 &  \\
314.3 & -10.24 & -12.46 & -10.18 & -11.03 &       & 	& -10.63 &        & -9.55 & -11.12 &  \\
319.3 & -10.19 & -12.40 & -10.16 & -10.94 &       & 	& -10.66 &        & -9.58 & -11.15 &  \\
326.3 & -10.13 & -12.38 & -10.12 & -11.08 &       & 	& -10.61 &        & -9.58 & -11.14 &  \\
338.4 & -10.50 & -12.71 & -10.43 & -11.42 &       & 	& -10.86 & -11.52 & -9.73  & -11.35 &  \\
341.4 & -10.26 & -12.60 & -10.29 & -11.24 &       & 	& -10.86 & -11.62 & -9.73  & -11.40 &  \\
346.4 & -10.44 & -12.63 & -10.36 & -11.20 &       & 	& -10.82 &        & -9.73  & -11.32 &  \\
352.4 & -10.29 & -12.44 & -10.27 & -11.09 &       & 	& -10.75 &        & -9.66  & -11.22 &  \\
366.4 & -10.47 & -12.57 & -10.48 & -11.80 &       & 	& -11.65 & -10.97 & -9.71  & -11.16 &  \\
382.4 & -10.42 & -12.59 & -10.45 & -11.21 &       & 	& -10.92 &        & -9.79  & -11.40 &  \\
387.4 & -10.46 & -12.67 & -10.46 & -11.22 &       & 	& -10.93 &        & -9.83  & -11.42 &  \\
395.4 & -10.60 & -12.85 & -10.61 & -12.09 &       & 	& -11.24 & -11.19 & -9.91  & -11.69 &  \\
424.5 & -10.58 & -12.84 & -10.70 & -12.04 &       & 	& -11.35 & -11.29 & -9.97  & -11.74 &  \\
463.5 &	-10.89 & -12.88	& -10.96 &	  &	  & 	& -11.73 & -10.89 & -10.21 & -12.11 &  \\
478.5 &	-10.76 & -12.97	& -10.88 &	  &	  & 	& -11.70 & -10.97 & -10.18 & -11.98 &  \\
507.4 & -10.62 & -12.88 & -10.81 &        &       & 	& -11.57 & -11.11 & -10.09 & -11.87 &  \\
555.4 &	-10.67 & -12.90	& -10.89 &	  &	  & 	& -11.70 & -11.20 & -10.18 & -12.08 &  \\
588.4 & -10.90 & -12.98 & -11.06 &        &       &     & -12.25 & -10.97 & -10.43 & -12.23 &  \\
614.4 & -11.09 & -13.26 & -11.27 &        &       &     & -12.29 & -11.10 & -10.57 & -12.66 &  \\
626.5 &	-11.24 & -13.23	& -11.38 &	  &	  &	& -12.35 & -11.08 & -10.66 &        &  \\
642.5 &	-11.21 & -13.27	& -11.36 &	  &	  &	& -12.40 & -11.16 & -10.66 & -12.56 &  \\
657.5 &	-11.31 & -13.37	& -11.52 &	  &       &     & -12.73 & -11.35 & -10.97 &        &  \\
662.5 & -11.34 & -13.31 & -11.49 &        &       &     & -12.80 & -11.16 & -10.72 & -12.46 &  \\
&&\\

	\hline
	\end{tabular}
	\end{table}

The rise to maximum brightness was primarily due to a tenfold increase in
the continuum, which can be reasonably fitted over the optical wavelength
range by an A-type pseudo photosphere passing in radius from R$\approx$35
R$_\odot$ to R$\approx$100 R$_\odot$.  The passage at maximum brightness was
characterized by FeII becoming prominent, together with reinforcement of HI
Balmer and Paschen series, while all HeI lines briefly vanished (cf. 
Figure~\ref{fig_dynamic_b}), well matching the Williams (1992) requirements
for a classification of NCas21 as a FeII-type nova.

From the spectroscopic point of view, the passage at primary maximum changed
irreversibly the profiles of emission lines: as illustrated in
Figure~\ref{fig_12_date}, prior to maximum the emission component displayed
a slim, symmetric Voigt profile, afterward the emission profile turned much
wider with rounded tops and a multi-component structure (a broad core
superimposed on an even broader pedestal), more than doubling the overall
width.  So, while from a photometric point of view the passage at maximum
has been just an isolated episode, with brightness and colors returning
in a week to pre-max values, it marked instead an irreversible change
for the spectra and the emersion of high-velocity components.  A sample of
the emission line profiles during the rise to maximum (day +49.7), passage
at (day +53.7), and return to pre-maximum brightness (day +61.5) are
presented in the 3rd, 4th, and 5th panels of Figure~\ref{fig_12_date}.

It was at this stage, when the high-velocity flow appeared in the spectra,
that for a 5-days interval following the return to pre-max brightness (days
+62 to +66), $\gamma$-rays from NCas21 were observed for the first time by
Fermi-LAT, as reported by Buson et al.  (2021).  The $\gamma$-rays detection
was preceded by persistent non-detections (Gong \& Li 2021), with an upper
limit of 1.5e34 erg/s for a 1.7kpc distance to NCas21.  It is worth noticing 
that shock-powered $\gamma$-rays emerged simultaneous with the drastic
transformation of the emission line profiles: the broad core superimposed on
an even broader pedestal indicates distinct large masses of gas moving at
greatly different velocities, conducive to powerful collisions and shocks. 
The $\gamma$-rays detection triggered radio observations with VLA, looking
for synchrotron signatures of particle-accelerating shocks, but none was
found in June, and only thermal emission from the nova was recorded by
Sokolovsky et al.  (2021b).

	\begin{figure}[!ht]
	\centering
	\includegraphics[width=15.5cm]{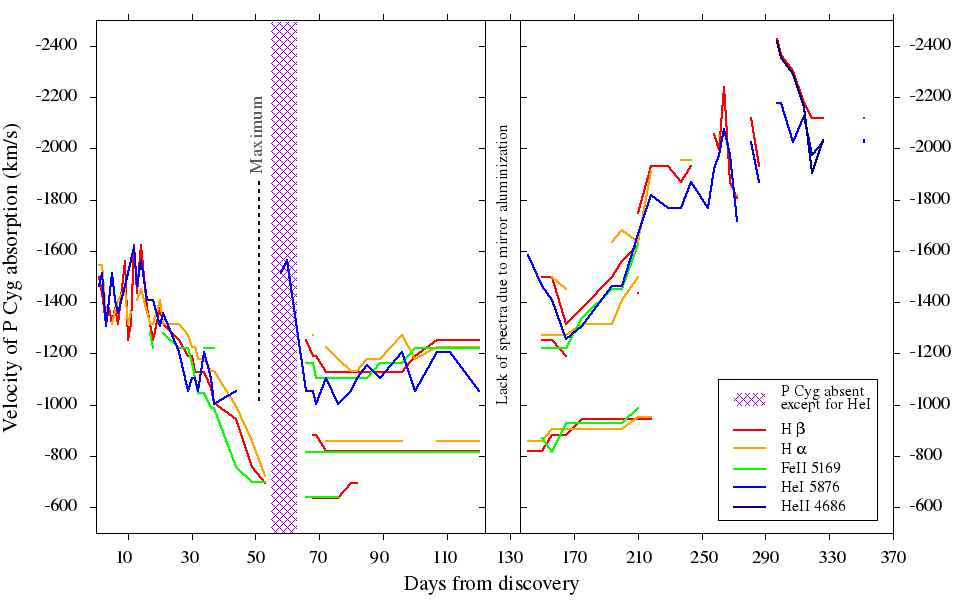}
	\caption{Heliocentric velocity of the P-Cyg absorption component(s)
	to H$\alpha$, H$\beta$, HeI 5876, FeII 5169, and HeII 4686 line
	profiles.  Absorptions belonging to the \emph{pre-maximum} system
	(following McLaughlin 1960 nomenclature) showed a rather turbulent
	behavior until day +15, after which their velocity smoothly declined
	from about $-$1400 to $-$700~km/s, and then disappeared just before
	the nova passing through maximum brightness (the purple strip marks
	the interval of time during which \emph{all} P-Cyg pre-maximum
	absorptions, temporarily disappeared from the spectrum of NCas21. 
	The gap between days 122-142 is instead due to lack of observations
	due to telescope refurbishment).  They only briefly persisted
	post-maximum, quickly replaced by multi-component absorptions
	belonging to the \emph{principal} system.  \emph{Principal}
	absorptions were composed of multiple sharp components at a fairly
	stable mean velocity of $\simeq 800$~km/s.  The \emph{diffuse
	enhanced} system of absorptions developed a broad component that
	steadily increased its velocity from $-$1200 to $-$2200~km/s, which
	after day +210 splitted into sub-components.  The time scale
	of the plot is compressed by a factor of two after day +130 to
	account for the slower evolution in the final stages.}
	\label{fig_P_Cyg} 
	\end{figure}

	\begin{figure}[!hb]
	 \begin{minipage}[b]{7.6cm}
	   \centering
	   \includegraphics[width=7.5cm]{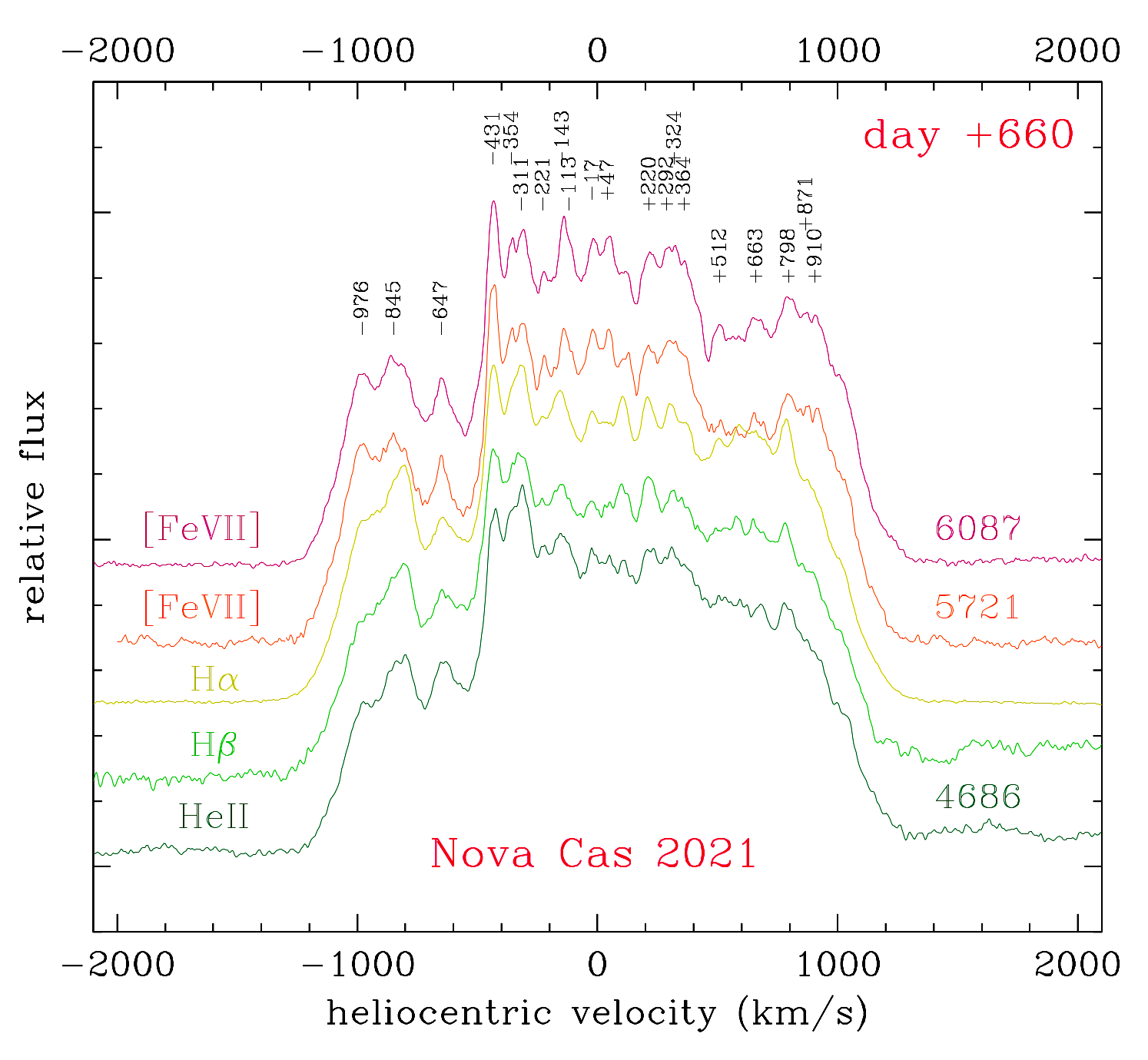}
	   \caption{Comparison in the velocity space of the castellated profiles
           of selected emission lines recorded on day +662 for Nova Cas 2021 .
           The heliocentric velocities (km/s) measured for [FeVII] 6087 are marked.} 
	\label{fig_profili_662}
	 \end{minipage}
	 \ \hspace{0.1mm} \hspace{0.2mm} \
	 \begin{minipage}[b]{8.5cm}
	  \centering
	   \includegraphics[width=8.4cm]{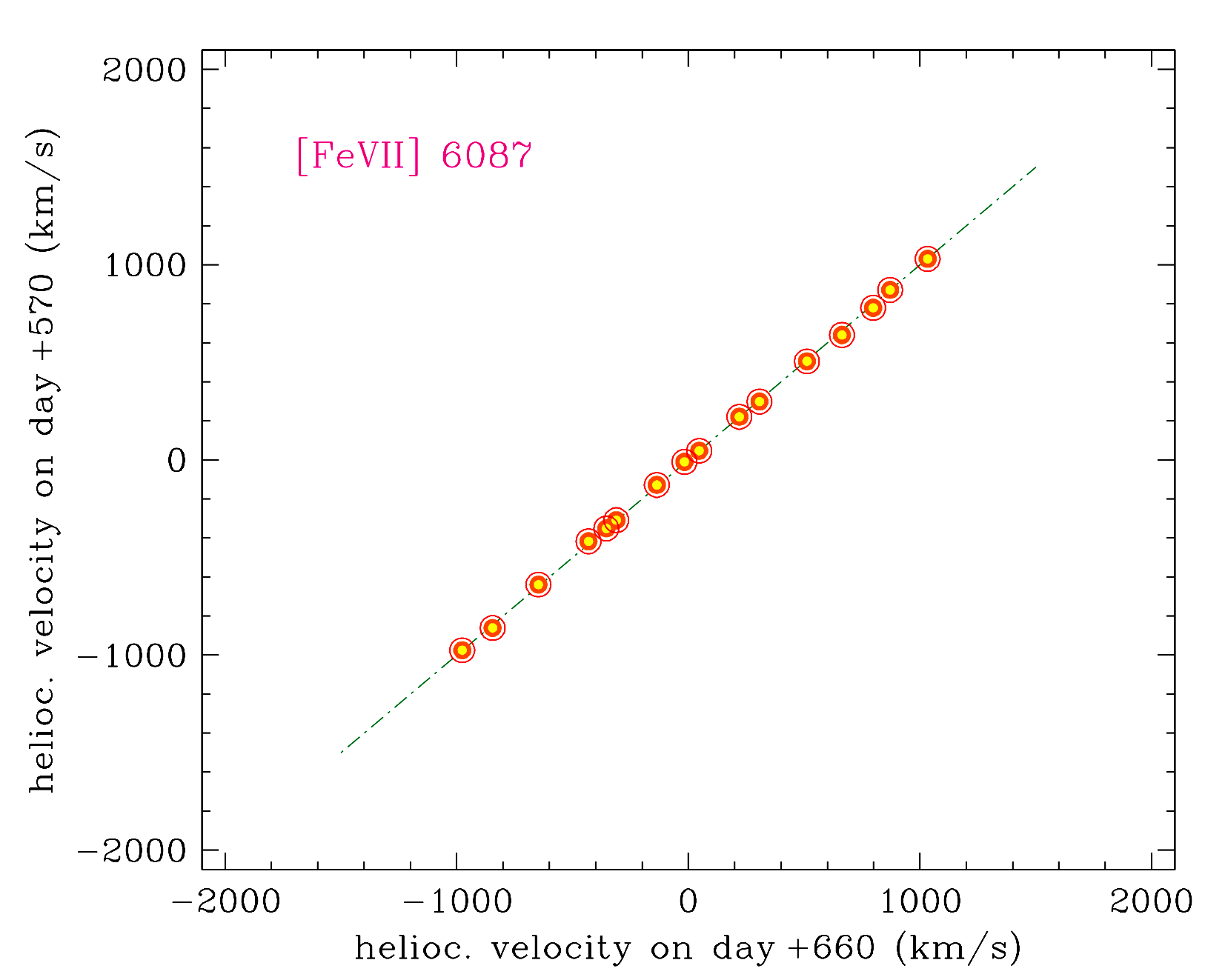}
	   \caption{Comparison of the heliocentric velocities of the features
           measured on the castellated profile of [FeVII] 6087 emission line on day +660
           (profile on the figure at left) and on day +570. The dot-dashed line
           represent the 1:1 relation.}
	\label{fig_deceleration}
	 \end{minipage}
	\end{figure}
\clearpage

The quick decline from maximum was characterized by the disappearance of all
absorptions components, the only weak absorptions remaining on day +61.5
being those of NaI D1,D2 doublet; on the same date, emission lines had red
and blue peaks at $+$650 and $-$815 km/sec, flanking asymmetrically a
brighter central component.  The FWZI of the emission lines was on this date
$\sim$2000~km/s.

Once the decline from maximum was completed, and the brightness had returned
to the plateau level, P-Cyg absorptions resumed for H$\beta$, Fe II and He I
lines, this time with the noticeable characteristics of being
multiple-component instead of single-component as before the primary
maximum.  A broad P-Cyg component at $-$1050 km/sec appeared after maximum
for HeI, FeII and H$\beta$, resembling the \textit{diffuse enhanced}
absorption system described by McLaughlin (1942), while the
\textit{principal} absorption system characterized by a $-$700 km/sec just
before maximum, splitted into two sharp components at $-$600 and $-$800
km/sec and re-appeared only for H$\beta$ and FeII but not for HeI (cf. 
Figures~\ref{fig_dynamic_b} and \ref{fig_12_date}).  The \textit{principal}
absorptions maintained a rather constant velocity along the transition
phase.  A summary of the evolution of P-Cyg absorptions is presented in
Figure~\ref{fig_P_Cyg} and is qualitatively consistent with the
time-behavior described by McLaughlin (1942), with \textit{pre-maximum}
speed decreasing before maximum and \textit{diffuse enhanced} speed
increasing after maximum, and reaching higher velocities than
\textit{principal} absorptions.

The empirical relation found by McLaughlin (1960) between the velocity of
principal absorption system and $t_3$ decline time 
\begin{equation}
\log {V_p} = 3.7 - 0.5\cdot \log {t_3}
\end{equation} 
provides $ t_3 \approx 50$ days for the average $V_p = 700$ km/sec presented
by NCas21, which is much shorter than any value of $t_3$ returned by
photometry as listed in Table~\ref{tab_t3}.  Rather interestingly, however,
inserting such \textit{spectroscopic} $ t_3 $ into the MMRD relation by
Selvelli \& Gilmozzi (2019), results in an absolute magnitude at maximum of
NCas21 of $M_V =-7.5 $ which leads to a distance of 1.8~kpc, in perfect
agreement with the 1.73 (range 1.66$\leftrightarrow$1.81) kpc value inferred
from the Gaia DR3 parallax.

\subsection{The plateau phase and secondary maxima}

During the 7-month photometric plateau, the spectrum showed prominent
Balmer, HeI and weaker FeII lines, with varying intensities, flanked by
variable and multiple P-Cyg absorptions due to \textit{principal} and
\textit{diffuse enhanced} systems.  Superimposed on the photometric plateau,
various rebrightenings have been observed, the brightest five being listed
in Table~\ref{tab_maxima}.  Thanks to the fast pace of our monitoring
campaign, we have recorded spectra covering all maxima and the in-between
returns to plateau brightness.

The spectra obtained around secondary maxima display some common spectroscopic
features. Comparing for ex. the spectra for day +142 (plateau) and +176 (N.3
maximum) in Figure~\ref{fig_12_date}, shows that at maxima the spectra
presented ($i$) lower ionization conditions with FeII reinforcing and HeI
weakening, ($ii$) stronger P-Cyg absorptions and a reduced velocity for them,
($iii$) narrower emission components, ($iv$) a reinforced continuum, and
($v$) a dramatic reduction of OI 8446 pumped by Lyman-$\beta$ fluorescence.

Aside from secondary maxima, the general trend exhibited instead by NCas21 along
the underlying plateau has been characterized by: \textit{(a)} stable
ionization conditions, higher than during maxima; \textit{(b)}
weakening of \textit{principal} system of P-Cyg absorptions with a constant
velocity for FeII and Balmer lines around $-$800~km/s, while a
\textit{principal} system of absorptions never developed for HeI;
\textit{(c)} the velocity of the \textit{diffuse enhanced} system of P-Cyg
absorptions increased from about $-$1100 to $-$1500~km/s; and \textit{(d)} a
weak auroral [NII] 5755 line was present from day +97 to +166 with a
temporary maximum intensity reached on day +142.  

The spectrum for day +142 is remarkable because it shows a temporary
and almost complete disappearance of \textit{diffuse enhanced} P-Cyg
absorptions for FeII and Balmer lines whose emission profiles
widened well beyond the \textit{principal} absorptions.
For HeI lines \textit{diffuse enhanced} P-Cyg absorption temporarily increased
to $-$1590~km/s velocity.

Also on this date, the H$\alpha$/H$\beta$ and OI 8446/7774 ratio reached 
the highest values for NCas21 (cf Figure~\ref{fig_dereddened}) and the high 
ionization HeII 4686 and Bowen 4640 blend turned briefly visible, to quickly 
disappear within a couple of weeks and then wait for the start of the decline 
in early November 2021 to return and remain there permanently visible.

\subsection{The nebular phase and photometric decline}

The photometric plateau of NCas21 ended around day +230, the rebrightenings
stopped altogether, and the nova finally embraced the steady decline it is
still following at the time of writing.

	\begin{figure}[!ht]
	\centering
	\includegraphics[width=16.5cm]{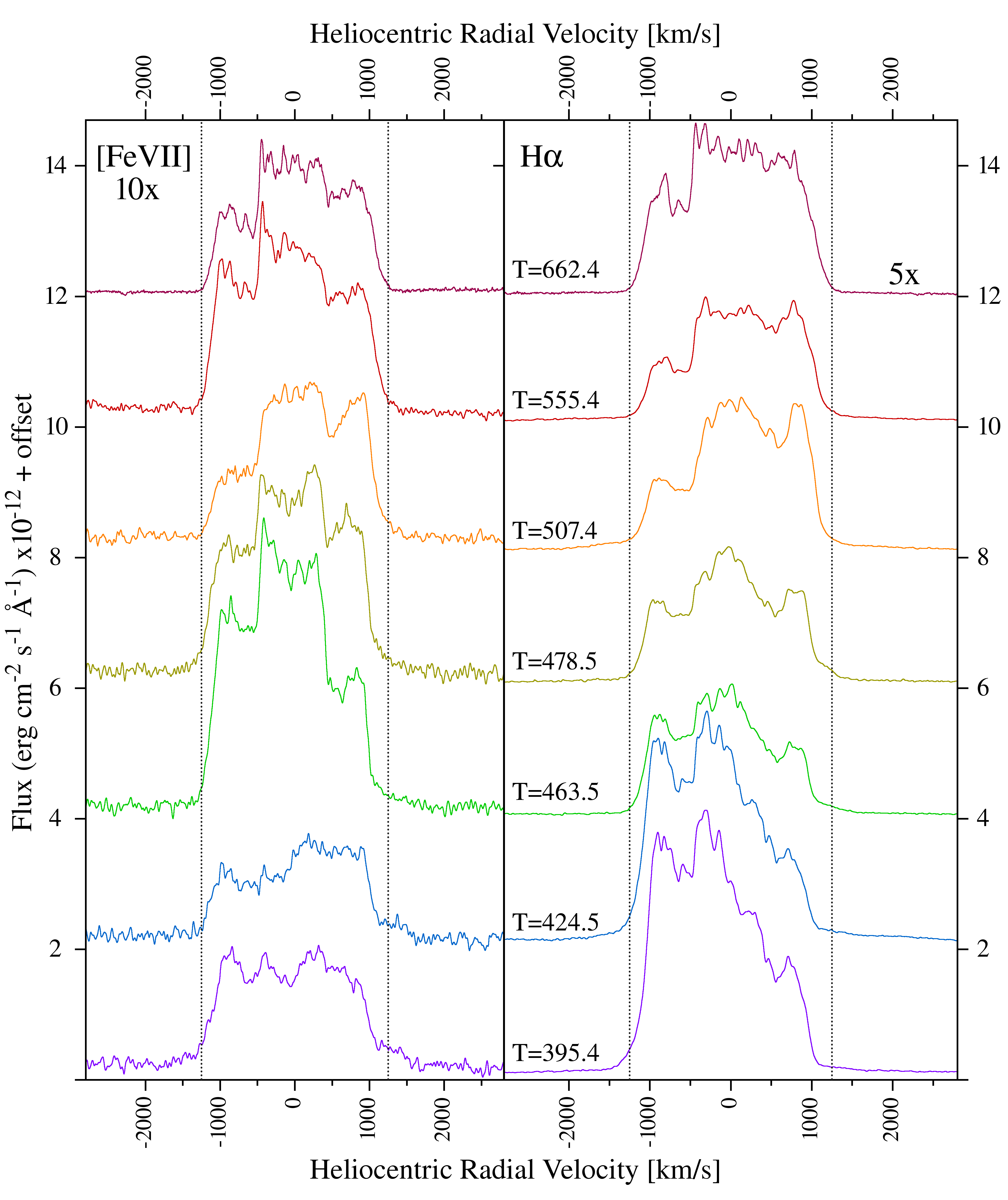}
	\caption{Comparison between the evolution of the emission line
        profile of [FeVII] 6087 and H$\alpha$ bracketing in time the VLA
        radio map of day +472 by Sokolovsky et al.  (2022b) that spatially
        resolved the radio shell around Nova Cas 2021.  Note how, since day
        +463, the profiles maintained a nested/multi-component boxy profile. 
        For [FeVII] on day +662 the trapezoidal pedestal is 2620~km/s wide
        at the base and 1930~km/s at the top, and a central core is 930~km/s
        wide at the base and 800~km/s at the top.}
	\label{Fig_Fe7_Ha}
	\end{figure}

	\begin{figure}
	\centering
	\includegraphics[width=16.5cm]{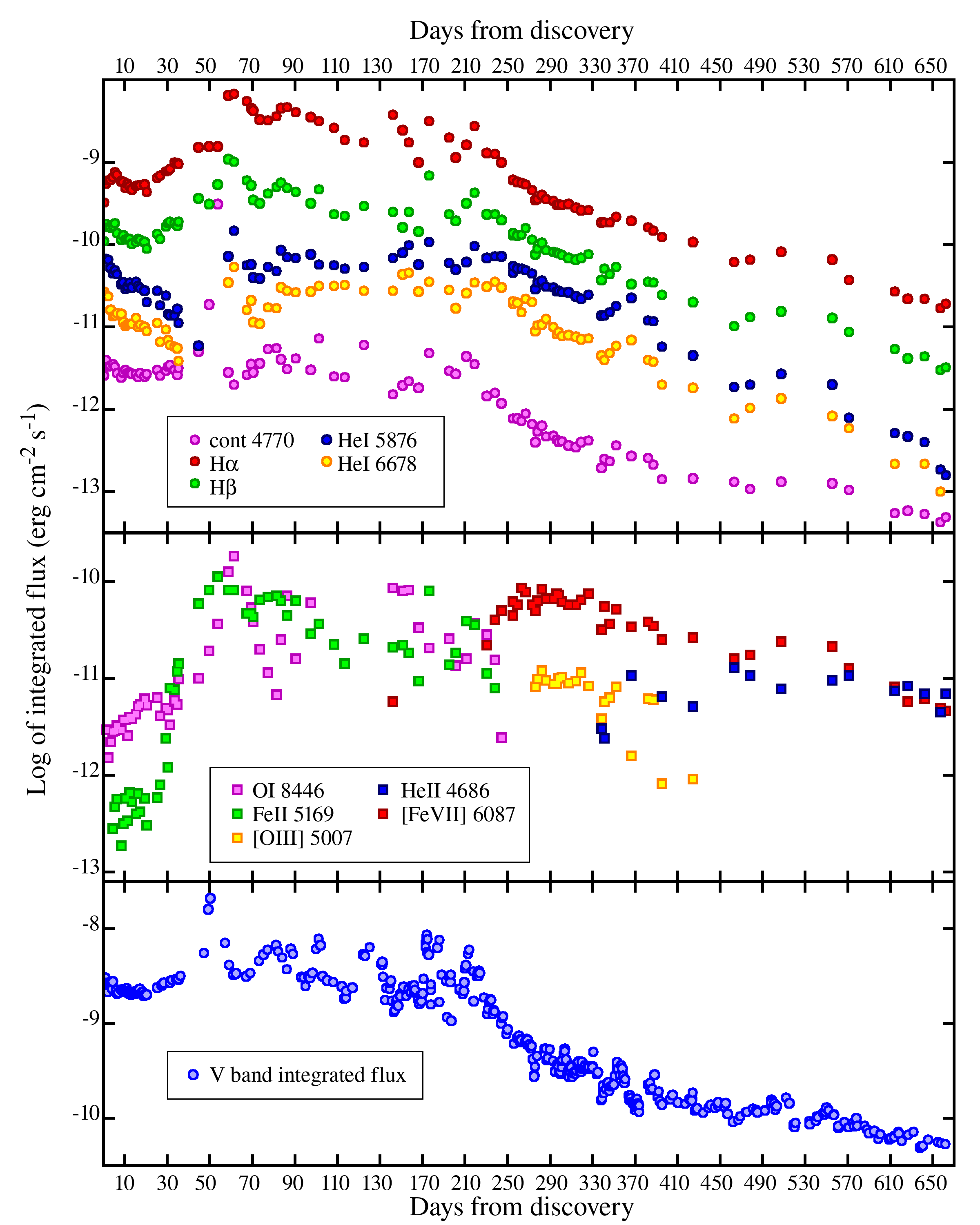}
	\caption{{\em Top panel}: evolution of the integrated flux of
	H$\alpha$, H$\beta$, He I 6678, and HeI 5876 emission lines (log
	scale) compared with the background continuum measured at 4770 \AA\
	(expressed in erg cm$^{-2}$ s$^{-1}$ \AA$^{-1}$), a region free from
	emission lines.  {\em Center panel}: evolution of FeII 5169, OI
	8446, HeII 4686, [OIII] 5007, and [FeVII] 6087.  {\em Bottom panel}:
	Integrated flux collected through the $V$ band.}
	\label{fig_line_fluxes}
	\end{figure}

	\begin{figure}
	\centering
	\includegraphics[width=16.5cm]{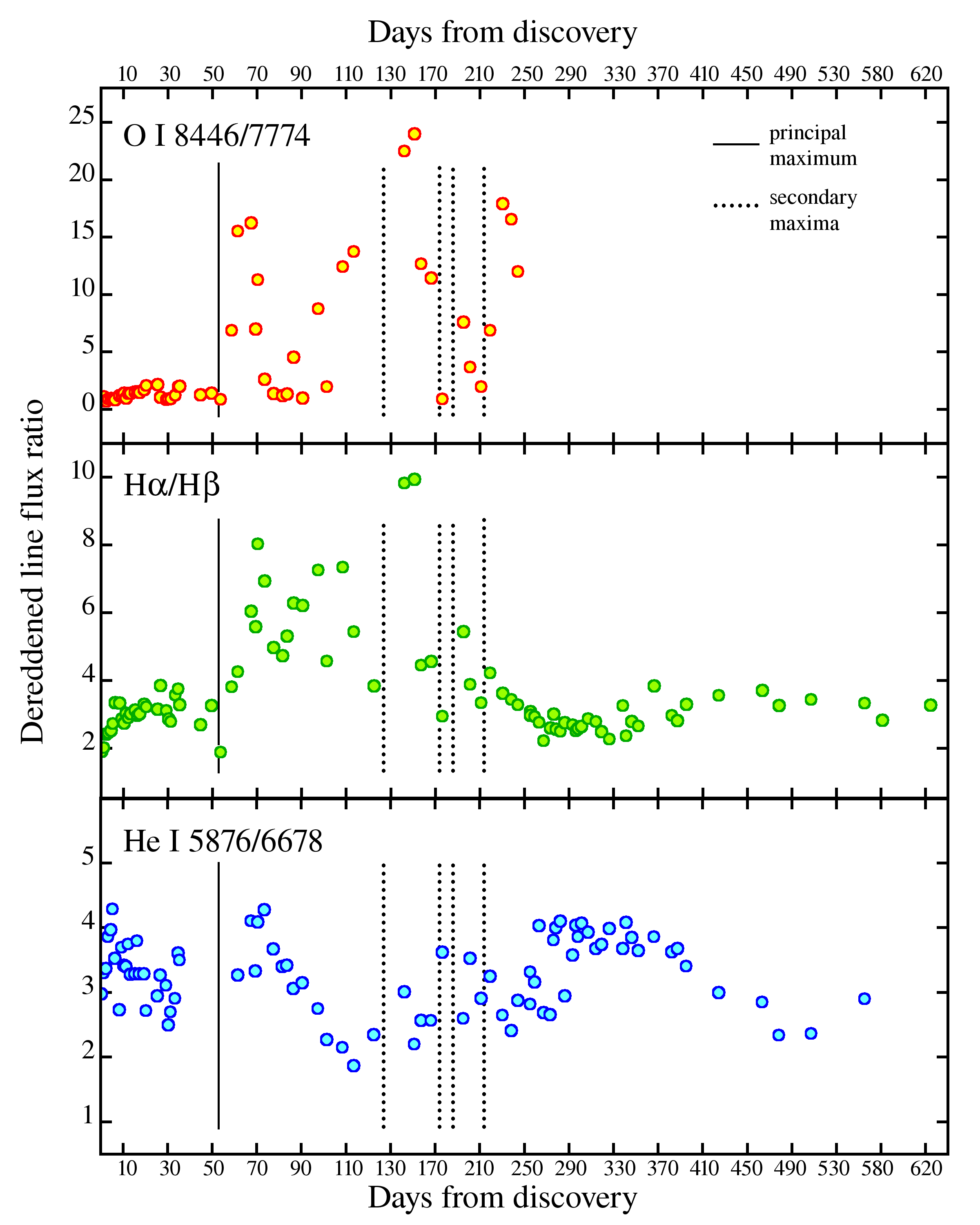}
	\caption{De-reddened line flux-ratios.  The solid vertical line mark
	the primary maximum, the dotted ones the secondary maxima listed in
	Table~\ref{tab_maxima}.  \textit{Top panel}: OI 8446/7774. 
	\textit{Center panel}: H$\alpha$/H$\beta$.  {\em Bottom panel}: HeI
	5876 triplet / 6678 singlet.  The flux of HeI 5876 line has been
	corrected for overlapping NaI and telluric absorptions, but some
	uncertainty affects such corrections for days +58 to +90.}
        \label{fig_dereddened}
	\end{figure}

\clearpage

During the final decline from the plateau, the ionization increased as
illustrated by the appearance and steady growth in intensity of HeII and of
the 4640 Bowen blend, while low-ionization features like FeII disappeared. 
The auroral [NII] 5755 soon appeared as the decline commenced.  Among other
slow novae, nebular stage was reported after 8 months for HR Del (Naito et
al.  2012), 18 months for V723 Cas (Iijima 2006), and 50 months for V1280
Sco (Naito et al.  2022).

The [NII] emission line in NCas21 displayed a boxy shape of FWZI$\sim$1900
km/s, with continuously varying blue and red edges; it reached maximum
intensity around day +255, and then disappeared by day +278 (see
Figure~\ref{fig_dynamic_b}).  With the start of the nebular phase, the P-Cyg
absorptions of Balmer and FeII lines rapidly declined while remaining strong
only for HeI up to day +320, the last time P-Cyg absorptions were visible
for any of the emission lines of NCas21.  From day +244 to +330 the velocity
of P-Cyg absorptions, although characterized by a steady overall increase in
velocity, displayed shorter term variability of a great amplitude (see
Figure~\ref{fig_P_Cyg}).

By day +274, [OIII] lines turned visible, in coincidence with the emergence
of super-soft X-rays emission (Page et al.  2021), and the rise in the
ionization conditions led to the disappearance of OI 7774 and 8446 lines.  
The last spectrum showing OI 8446 in emission is that for day +244.  

Higher ionization lines [FeVI], [FeVII] and [CaV] begun to appear on day
+320, initially with an intermittent intensity, anti-correlating with the
strength of [OIII] lines.  Both variations in the temperature of the
ionizing radiation field and in the electron density of the medium may have
contributed to such a behavior.  The measured [OIII] 5007/4959 ratio has
consistently remained above the theoretical value of 3, probably driven by the
blending of [OIII] 5007 with HeI 5016 line.  [OIII] 5007 finally disappeared
altogether around day +450; it was never very strong though, peaking in
intensity on day +319 when it reached only 1/8 of the H$\beta$ integrated
flux (cf.  Table~\ref{tab_log}).  Since about day +250 (cf. 
Figure~\ref{fig_12_date}), the Balmer and He I lines developed a
nested/multi-component profile, with a box-shaped central core
(FWHM$\simeq$800 km/s), superimposed onto a trapezoidal
FWHM$\simeq$2000~km/s base, and a final wide pedestal of FWZI$\simeq$4200
km/s.  The latter, on later dates and for H$\alpha$ reached a
FWZI$\simeq$5300 km/s.  The nested/multi-component boxy profile was
successively developed also by [FeVI], [FeVII], and [CaV] lines.

A spatially resolved radio map of the ejecta was obtained with VLA on
2022-07-04 (day +472) by Sokolovsky et al.  (2022b).  Their 31.1 and 34.9
GHz images show an elliptical edge-brightened shell extending about 220 mas
in the north-east to south-west direction, crossed by a bright band of
emission which extends along the minor axis of the shell.  The
Figure~\ref{Fig_Fe7_Ha} illustrates the evolution of the profile of
H$\alpha$ and [FeVII] 6087 of NCas21 around the epoch of VLA imaging of the
resolved ejecta, from 77-days before to 83-days after; they are
representative of permitted and forbidden emission lines.  At day +507 (35
days past the VLA epoch) the line profiles of both H$\alpha$ and [FeVII]
were still changing with the red and the blue wing of the pedestal
alternating as the stronger of the two.  It is only with day +555 (83 days
past the VLA epoch) that the emission line profiles stopped evolving and
remained freezed in their appearance.  This is well illustrated in
Figure~\ref{fig_profili_662}, where the densely castellated profiles of
emission lines for day +570 are plotted (left panel), and the velocity of
each dent compared to that measured on our last spectrum for day +662 (right
panel): the velocities at the two epochs are identical within the
measurement error (about $\pm$1 km/s), proving the ejecta had finally
reached the stage of free-expansion into the void surrounding the central
nova.

\bigskip
\noindent
{\sc \large Acknowledgements} We express our gratitude to L. Buzzi, A.
Frigo, S. Moretti, F. Castellani, V. Andreoli, and A. Bergamini for their
various help and support to this project.

\bigskip

\section{References}

\phantom{.}

Bhatia, A.~K., Kastner, S.~O., 1995, ApJS, 96, 325

Bruch, A., 1982, PASP, 94, 916 

Buson, S., Cheung, C.~C., Jean, P., 2021, ATel, 14658

Chomiuk, L., Linford, J.~D., Aydi, E., et al., 2021, ApJS, 257, 49 

Gehrz, R.~D., Banerjee, D.~P.~K., Evans, A., et a., 2021, ATel 14794

Gong, Y.-H., Li, K.-L., 2021, ATel, 14620 

Goranskij, V.~P., Katysheva, N.~A., Kusakin, A.~V., et al., 2007, AstBu, 62, 125 

Harman, D.~J., O'Brien, T.~J., 2003, MNRAS, 344, 1219 

Henden, A., Munari, U., 2014, CoSka, 43, 518 

Heywood, I., O'Brien, T.~J., Eyres, S.~P.~S., et al.,  2005, MNRAS, 362, 469 

Iijima, T., 2006, A\&A, 451, 563 

Landolt, A.~U., 2009, AJ, 137, 4186  

Lyke, J.~E., Campbell, R.~D., 2009, AJ, 138, 1090 

Maehara, H., Taguchi, K., Tampo, Y., et al., 2021, ATel, 14471 

McLaughlin, D.~B., 1942, ApJ, 95, 428

McLaughlin, D.~B., 1960, in Stellar Atmospheres, ed J.L. Greenstein (U. of Chicago Press), p.585

Munari, U., 2014, ASPCS, 490, 183

Munari, U., Moretti, S., 2012, BaltA, 21, 22 

Munari, U., Valisa, P., 2014, CoSka, 43, 174

Munari, U., Valisa, P., 2022, ATel, 15796

Munari, U., Zwitter, T., 1997, A\&A, 318, 269

Munari, U., Goranskij, V.~P., Popova, A.~A., et al., 1996, A\&A, 315, 166 

Munari, U., Bacci, S., Baldinelli, L., et al., 2012, BaltA, 21, 13

Munari, U., Henden, A., Frigo, A., et al., 2014, AJ 148, 81  

Munari, U., Maitan, A., Moretti, S., et al., 2015, NewA, 40, 28

Munari, U., Valisa, P., Dallaporta, S., 2021a, ATel, 14476 

Munari, U., Valisa, P., Dallaporta, S., 2021b, ATel, 15614

Munari, U., Valisa, P., Dallaporta, S., et al., 2021c, ATel, 15093

Naito, H., Mizoguchi, S., Arai, A., et al., 2012, A\&A, 543, A86

Naito, H., Tajitsu, A. , Ribeiro, V.~A.~R.~M., et al.,  2022, ApJ, 932, 39

Nayana, A.~J., Anupama, G.~C., Banerjee, D.~P.~K., et al., 2022, ATel, 15383

Ochner, P., Moschini, F., Munari, U., et al., 2015, MNRAS, 454, 123 

Page, K.~L., Starrfield, S., Munari, U., et al., 2021, ATel, 15111

Rafanelli, P., Rosino, L., 1978, A\&AS, 31, 337

Rudy, R., Subasavage, J., Bayless, A., et al., 2021, ATel, 14482

Schwarz, G.~J., Ness, J.-U., Osborne, J.~P., et al., 2011, ApJS, 197, 31  

Selvelli, P., Gilmozzi, R., 2019, A\&A, 622, A186 

Shore, S.~N., Teyssier, F., Garde, O., et al., 2021a, ATel, 14622 

Shore, S.~N., Buil, C., Dubovsky, P., et al.,  2021b, ATel, 15577

Sokolovsky, K.~V., Aydi, E., Chomiuk, L., et al., 2021a, ATel, 14530 

Sokolovsky, K.~V., Aydi, E., Chomiuk, L., et al., 2021b, ATel, 14731

Sokolovsky, K.~V., Aydi, E., Chomiuk, L., et al., 2022a, ATel, 15150

Sokolovsky, K.~V., Aydi, E., Chomiuk, L., et al., 2022b, ATel, 15518

Taguchi, K., Maehara, H., Isogai, K., et al., 2021a, ATel, 14472

Taguchi, K., Isogai, K., Shibata, M., et al., 2021b, ATel, 14478 

Van den Bergh, S., Younger, P.~F., 1987, A\&AS, 70, 125 

Warner, B., 1995, Cataclysmic Variable Stars, Cambridge Univ. Press.

Williams, R.~E., 1992, ApJ, 104, 2 

Woodward, C.~E., Banerjee, D.~P.~K., Evans, A., 2021, ATel, 14665

\end{document}